\documentclass{aa}
\usepackage[varg]{txfonts}
\usepackage[english]{babel}
\usepackage{amsmath}
\usepackage{graphicx}
\usepackage{capt-of}
\usepackage{url}

\begin{document}

\title{Protoplanetary disk masses in NGC 2024: Evidence for two populations}

\author{S.E.~van Terwisga\inst{\ref{leiden1},\ref{mpia}}\and E.F. van Dishoeck\inst{\ref{leiden1},\ref{garching}}\and R. K. Mann\inst{\ref{nrc}} \and J. Di Francesco\inst{\ref{nrc}} \and N. van der Marel\inst{\ref{nrc}}\and M. Meyer\inst{\ref{umich}}\and S.M. Andrews\inst{\ref{cfa}} J. Carpenter\inst{\ref{jao}}\and J.A. Eisner\inst{\ref{uaz}}\and C.F. Manara\inst{\ref{eso}}\and J.P. Williams\inst{\ref{hawaii}}}

\institute{
	Leiden Observatory, Leiden University, PO Box 9513, 2300 RA Leiden, The Netherlands\label{leiden1}\and
	Max-Planck-Institut f{\"u}r Astronomie, K{\"o}nigstuhl 17, 69117 Heidelberg, Germany\label{mpia}~\email{terwisga@mpia.de}\and
	Max-Planck-Institut f{\"u}r Extraterrestrische Physik, Gie{\ss}enbachstraße, D-85741 Garching bei M{\"u}nchen, Germany\label{garching}\and
	NRC Herzberg Astronomy \& Astrophysics, 5071 W Saanich Rd, Victoria BC V9E 2E7, Canada\label{nrc}\and
	Department of Astronomy, University of Michigan, 1085 S. University, Ann Arbor, MI 48109, USA\label{umich}\and
	Harvard-Smithsonian Center for Astrophysics, 60 Garden Street, Cambridge, MA 02138, USA\label{cfa}\and
	Joint ALMA Observatory, Av. Alonso de C\'{o}rdova 3107, Vitacura, Santiago, Chile\label{jao}\and
	Steward Observatory, University of Arizona, 933 North Cherry Avenue, Tucson, AZ 85721, USA\label{uaz}\and
	European Southern Observatory, Karl-Schwarzschild-Str. 2, D-85748 Garching bei M{\"u}nchen, Germany\label{eso}\and
	Institute for Astronomy, University of Hawai`i at M{\=a}noa, 2680 Woodlawn Dr., Honolulu, HI, USA\label{hawaii}
	%Max-Planck-Institut f{\"u}r Extraterrestrische Physik, Gie{\ss}enbachstraße, D-85741 Garching bei M{\"u}nchen, Germany\label{garching}
}

\abstract{Protoplanetary disks in dense, massive star-forming regions  are strongly affected by their environment. How this environmental impact changes over time is an important constraint on disk evolution and external photoevaporation models.}
{We characterize the dust emission from 179 disks in the core of the young (0.5\,Myr) NGC 2024 cluster. By studying how the disk mass varies within the cluster, and comparing these disks to those in other regions, we aim to determine how external photoevaporation influences disk properties over time.}
{Using the Atacama Large Millimeter/submillimeter Array, a $2.9' \times 2.9'$ mosaic centered on NGC 2024 FIR 3 was observed at 225\,GHz with a resolution of $0.25''$, or $\sim100$\,AU. The imaged region contains 179 disks identified at IR wavelengths, seven new disk candidates, and several protostars.}{The overall detection rate of disks is $32 \pm 4\%$. Few of the disks are resolved, with the exception of a giant ($R = 300$\,AU) transition disk. Serendipitously, we observe a millimeter flare from an X-ray bright young stellar object (YSO), and resolve continuum emission from a Class 0 YSO in the FIR 3 core. Two distinct disk populations are present: a more massive one in the east, along the dense molecular ridge hosting the FIR 1-5 YSOs, with a detection rate of $45 \pm 7\%$. In the western population, towards IRS 1, only $15 \pm 4\%$ of disks are detected.}
{NGC 2024 hosts two distinct disk populations. Disks along the dense molecular ridge are young (0.2--0.5\,Myr) and partly shielded from the far ultraviolet radiation of IRS 2b; their masses are similar to isolated 1--3\,Myr old SFRs. The western population is older and at lower extinctions, and may be affected by external photoevaporation from both IRS 1 and IRS 2b. However, it is possible these disks had lower masses to begin with.}


\maketitle

\section{Introduction}
Protoplanetary disks are formed and evolve in a wide variety of environments: from low-mass, isolated, star-forming regions (SFRs) like the Taurus and Lupus clouds, to dense, massive SFRs like the Orion Nebula. Disks are not completely independent of their environment. For instance, IM Lup's extended CO halo has been interpreted as evidence of external photoevaporation~\citep{haworth17}. It is, however, in the massive star-forming regions that the impact of the environment on the evolution of disks becomes most pronounced. The proplyds in the Orion Nebula Cluster (ONC) have long been recognized as disks that are being ionized by UV radiation from the bright young stars in the Trapezium, losing mass as a result~\citep{odell93, odell94, mann14, eisner18}. Not only is the effect of environment on disks larger in massive star-forming regions, but more stars are formed in clusters overall~\citep[e.g.,][]{lada03, carpenter00, porras03}. Understanding the impact that these environments have on the evolution of disks, and in particular on the amount of mass available for planet formation as a function of time, is therefore important for our understanding of the observed planet population.

Surveys of individual populations of protoplanetary disks using their millimeter-continuum emission have, in recent years, become a key tool for studying disk evolution. The Atacama Large Millimeter/submillimeter Array (ALMA) has not just the sensitivity but also the resolution to resolve disks out to the distance of the Orion clouds. As a result, the continuum emission from cold, millimeter-sized dust grains in the disk is now an easily accessible observable. Disk (dust) mass distributions that are based on the assumption that continuum flux is proportional to disk mass are now available for most nearby low-mass star-forming regions~\citep[e.g.,][]{ansdell16,pascucci16,barenfeld16,ruizrodrigues18,williams19,cazzoletti19}.

In Orion, several areas of massive star formation can provide a counterpoint to the now well-studied low-mass environments. The ONC, in Orion A, is the richest cluster within 500 pc of the sun, with an age of about 1\,Myr. It has now been studied quite extensively with ALMA, revealing a strongly photoevaporated population of protoplanetary disks in the inner 0.5 pc around the massive stars in the Trapezium driven by the O6 star $\theta^{1}$ Ori C~\citep[e.g.,][]{mann14,eisner18}. Beyond 0.5\,pc of this star, however, the disk mass distribution is surprisingly close to that in Lupus and Taurus \citep{vanterwisga19}, which have comparable ages. The $\sigma$ Orionis region in Orion B is somewhat older at 3-5 Myr~\citep{oliveira02,oliveira04}, and its ionizing star ($\sigma$ Ori) is an O9 star, cooler than the Trapezium's most massive stars. Disks in this region, however, likewise show evidence of external photoevaporation in submillimeter observations~\citep{ansdell17}. In $\sigma$ Orionis, the radius out to which this effect is important is larger (~2 pc) than in the ONC, but its overall impact on the disk masses in the region is lower.

The protoplanetary disks of NGC 2024 are, in this context, an important missing link. NGC 2024 hosts the youngest population of young stellar objects (YSOs) in the Orion clouds, at 0.5\,Myr~\citep{meyer96,levine06}. While precise ages are difficult to determine, NGC 2024's stellar population is certainly young relative to the ONC~\citep{eisner03}. There is, however, evidence of a radial age gradient~\citep{getman14}. Apart from its youth, NGC 2024 is also the richest SFR in Orion B~\citep{meyer08}. There is an ongoing debate on the identification of the ionizing source(s) in the region. Unlike in the ONC, the source is not directly visible.~\citet{burgh12} find evidence of an O6 to B0-type star in the deeply extincted cluster core. IRS 2b is a candidate~\citep{bik03,kandori07}, but its spectral type may be too late. IRS 2b, IRS 2, and other possible sources located close to each other on the sky are the most likely drivers of ionization in the nebula~\citep{bik03, lenorzer04, meyer08}. Even outside the cluster's inner regions, IRS 1 (B0.5) may also contribute to the total flux~\citep{burgh12}.

With its young population and massive stars, NGC 2024 provides a perspective on how disks are affected by external UV irradiation in the first few millions of years after the envelope has dissipated. Previous observations of disks in NGC 2024 with (sub)millimeter interferometers have managed to detect a number of disks in this region, but did not resolve them, and are likely more contaminated by the complicated large-scale cloud emission in this SFR~\citep{eisner03,mann15}.

In this article, we use a large-scale ($2.9' \times 2.9'$) ALMA mosaic of the inner part of NGC 2024, centered on the dense molecular ridge~\citep{watanabe08}, and including IRS-2b and IRS 1 in the field. Our sample of 186 disks is larger than previous studies, while ALMA's sensitivity enables us to detect disks down to less than $1\,M_{\oplus}$ in dust. These observations cover a wide range of physical conditions, from the direct vicinity of IRS 2b, to the deeply obscured parts of the dense molecular ridge, to the less obscured environment to the west of the cluster core. Combined with the large sample size, this variety means that we can study how disk properties differ across the cluster. By comparing the dust mass distribution to that of disks in the ONC and nearby low-mass star-forming regions, we are able to trace the disk mass evolution across time in a photoevaporating sample.

\section{ALMA observations and data reduction}
The observations published in this article were taken as part of the ALMA program 2017.1.01102.S (PI: R. Mann), and consist of a Band 6 mosaic of a $2.9' \times 2.9'$ square region. The imaged area covers both the immediate environment of IRS 2b, the dense molecular ridge hosting the FIR-3 and FIR-2 sources, the FIR-4 source (resolved as a binary Class 0 protostar) and a relatively empty region to the west.

The full mosaic consists of 149 pointings, centered on $05^\mathrm{h}41^\mathrm{m}40.5^\mathrm{s}$, $-01^\circ54{}^\prime16.25{}^{\prime\prime}$ (J2000). Over the course of the observations, the full set of 149 pointings was observed on eight separate occasions. All observations used the same calibrators: amplitude and bandpass calibration were carried out on J0432-0120 and phase calibration was performed on J0541-0211.

The spectral setup of the observations covers the millimeter-continuum with two spectral windows centered at 230\,GHz and 219\,GHz. The effective rest frequency of the combined observations is 225\,GHz. Additionally, three spectral windows with 122\,kHz-wide channels cover the $J=2-1$ transitions of $^{12}$CO (with 1920 channels), $^{13}$CO, and C$^{18}$O (both with 960 channels). Due to the dense interstellar environment towards NGC 2024, however, in most pointings line emission from the cloud severely contaminates the science targets, or becomes optically thick and resolves out. Therefore, we focus here on the results from the continuum observations. The images presented here have been made after flagging the channels with strong line emission.

Images of the data were made in \texttt{CASA} 5.4.0, using the \textit{tclean} task. In all instances except where explicitly noted, we used Briggs weighting, with a robust parameter of $1.0$, for the best image signal-to-noise ratio (SNR) and a compact beam shape. The data span baselines from $10 - 1700$\,k$\lambda$. The observing schedule combined with the short integrations of the individual pointings for these observations led to the shortest baselines being sampled with very little field rotation. To suppress the resulting partially resolved-out emission from the dense molecular ridge, and to detect the (compact) disks more clearly, another image was generated where only baselines $>150$\,k$\lambda$ were included. The effective beam shape is $0.26^{\prime\prime} \times 0.18^{\prime\prime}$ (full width at half maximum) for the full baseline image. For the long-baseline image, the effective beam shape is $0.22^{\prime\prime} \times 0.15^{\prime\prime}$.

The data were self-calibrated to improve their SNR. Each integration of the mosaic was separately self-calibrated for phase only. The first full visit of the mosaic was affected by slightly higher noise levels and contains an extremely bright outburst from a variable object (see~Sect. \ref{sec:flare}). For this reason, the data from this first integration were not used for creating the images from which we derive the dust masses of disks. This exclusion does not significantly affect the noise level in the final images. Due to overlapping fields of slightly different depths, the full synthesized image has spatially variable noise. In the most sensitive part of the image, the rms noise is 0.052\,mJy\,beam$^{-1}$. After restricting the baseline coverage to baselines $>150$\,k$\lambda$, the noise increased to 0.058\,mJy\,beam$^{-1}$. The spatial variability in the noise level was taken into account during source extraction by weighting the noise estimated in an empty part of the image with the primary-beam coverage, as calculated by \textit{tclean}. This correction had a small effect ($<5\%$) everywhere but on the edges of the primary beam of the outermost pointings.

\section{Results}
\begin{figure*}[t]
        \begin{center}
                \includegraphics[width=\textwidth]{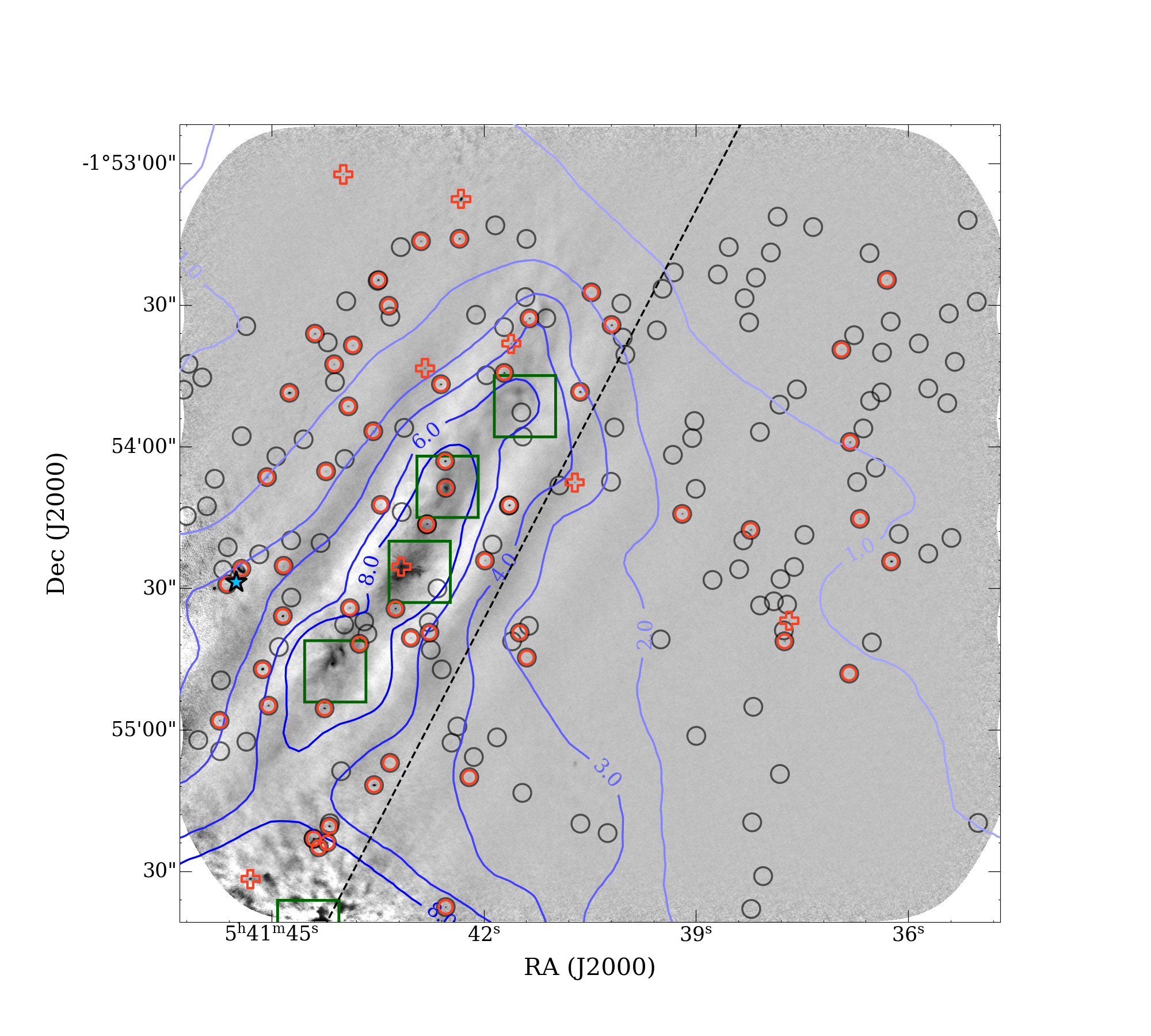}
                \caption{ALMA mosaic of the NGC 2024 core region at 225\,GHz using the full baseline coverage, in grayscale in the background. Black circles indicate the positions of the disks in the~\citet{meyer96} catalog; red circles show the detected sources. The open red plus symbols indicate the locations of additional point sources not included in the catalog. The dark blue star symbol marks the position of IRS 2b. The eastern and western populations (Section~\ref{sec:internal}) are separated by the thin black dashed line. Blue-shaded contours show the location of the dense molecular ridge containing the FIR 1 - 5 sources using {\it Herschel} PACS 160\,$\mu$m data in units of Jy\,pixel$^{-1}$), from~\citet{stutz13}. The locations of FIR 1 - 5 are marked (from north to south) by green squares.}
                \label{fig:disksinspace}
        \end{center}
\end{figure*}
In this section we present the results of our millimeter continuum survey of protoplanetary disks in NGC2024. For the detected sources, we derive dust masses and the disks' dust mass distribution. Several sources in the image are particularly interesting and discussed separately.

\subsection{A 1.3\,mm catalog of disks in NGC 2024}
To study the demographics of NGC 2024 disks with ALMA, it is essential to have a properly defined sample to study. In this paper, we used the catalog of infrared-excess objects from~\citet{meyer96}, which were identified by their JHK colors. Figure~\ref{fig:disksinspace} shows the ALMA observations (using all baselines) and overlays the catalog sources, marking the detections, to show the distribution of disks over the imaged area. The~\citet{meyer96} catalog is the largest in the field covered by the ALMA observations, ensuring an excellent sample size of 179 objects. Drawing the sample from this catalog means the selection criteria for sources are uniform. The excess emission at near infrared (NIR) wavelengths is due to hot, optically thick dust, and therefore should be a good tracer of disk presence without being biased to the mass of millimeter-sized grains in the cold midplane regions to which ALMA is sensitive. Only the extreme north of the field (the topmost $18''$ of the image in Fig.~\ref{fig:disksinspace}) is not covered in the catalog, leading to the exclusion of one bright, somewhat resolved source.

A significant limitation of the catalog used here is that, with only JHK-band photometry, it is possible that some objects may be incorrectly included. In particular, outflow cavities of younger, more-embedded objects may be mistaken for disks. While the spectral energy distributions (SEDs) of many sources are not well-sampled, we cross-referenced the~\citet{meyer96} catalog with the Herschel Orion Protostars Survey (HOPS) catalog of protostars in Orion~\citep{furlan16}. This comparison led us to exclude HOPS 384 (IRC 227 / 229) as a Class 0 source, but no other sources were rejected from the catalog for this reason. We also established that an embedded YSO that is associated with NGC 2024 FIR 3 in our ALMA observations is not detected in the~\citet{meyer96} catalog, suggesting that the misclassification of younger sources as disks should be rare. Finally, we note with~\citet{mann15} that JHK colors of the catalog members are generally not very red, indicating that most of them are dominated by disks and therefore representative of Class II disks.

For all sources in the base catalog, we performed aperture photometry on the brightest point source with emission $>3\,\sigma$ in the long-baseline image that was found within $1''$ of the catalog position, in order to account for astrometric uncertainties in the catalog. If no source was found, the flux was measured in a $0.25''$ beam on the catalog position. Detected source positions show no significant offset from the catalog positions on average. The resulting millimeter catalog was checked manually, and tested for the presence of extended emission in the full-baseline image. This last step was only necessary for the most radially extended disk in the field after IRS-2b. One source (IRC 115) was found to be a binary object, and has been split here into IRC 115 A and IRC 115 B.

Of the 179 objects in the sample, 57 are detected, with an overall detection rate of $32 \pm 4\%$. The brightest source in the field, IRC 101, has a flux of $204 \pm 1.1$\,mJy. The median flux for detected sources is 2.5\,mJy, while the faintest detected object has a flux of only $0.48 \pm 0.13$\,mJy. These errors do not include the standard (absolute) flux calibration accuracy of $10\%$. Zoomed-in cutouts of the detected sources are shown in Appendix~\ref{app:cutouts}, in Figs.~\ref{fig:cutouts1} and~\ref{fig:cutouts2}. The fluxes of the detected sources are listed in Table~\ref{tab:D}; the upper limits can be found in Table~\ref{tab:uplimmass} in Appendix~\ref{app:uplimmass}. Sources are detected throughout the field, as Fig.~\ref{fig:disksinspace} shows, even in regions with significant contamination from partly resolved-out cloud emission in the full-baseline image.

Our observations are primarily intended to detect, not to resolve, disks, and therefore the number of sources with resolved continuum emission is low. Using the \textit{imfit} task in \texttt{CASA}, we fit two-dimensional Gaussians to the sources in the image plane, and find meaningful results for the deconvolved source properties for the seven brightest disk-bearing sources (including IRS 2), which are presented in Table~\ref{tab:resolved}. The radii quoted for these sources are defined to enclose $98\%$ of the flux, assuming a Gaussian radial intensity distribution. IRC 101 is resolved, but for this source a two-dimensional Gaussian is not appropriate; we therefore provide manual estimates of the disk parameters. However, we caution that the second- and third-largest sources in this sample are also the most inclined, suggesting that our observations may miss faint extended emission around the bright inner disks.

\begin{table*}
\caption{Continuum fluxes and masses for the detected disks.}
\label{tab:D}
\centering
\begin{tabular}{l l l l l}
\hline \hline
Name & RA & Dec & Flux & Mass \\
 &  &  & mJy & $M_{\oplus}$ \\
\hline
IRS 2\tablefootmark{a}  & 5:41:45.81 & -1:54:30.0 & $241 \pm 32$ & - \\
IRC101 & 5:41:45.43 & -1:54:25.9 & $204.8 \pm 1.15$ & $1055.2 \pm 5.94$ \\
IRC067 & 5:41:45.13 & -1:54:47.2 & $37.1 \pm 0.18$ & $190.9 \pm 0.91$ \\
IRC086 & 5:41:44.84 & -1:54:35.9 & $27.0 \pm 0.39$ & $139.4 \pm 2.02$ \\
IRC044 & 5:41:43.55 & -1:55:11.8 & $18.5 \pm 0.13$ & $95.3 \pm 0.69$ \\
IRC153 & 5:41:44.75 & -1:53:48.6 & $15.3 \pm 0.39$ & $78.9 \pm 2.00$ \\
IRC215 & 5:41:36.24 & -1:54:24.4 & $13.0 \pm 0.15$ & $67.2 \pm 0.77$ \\
IRC036 & 5:41:44.18 & -1:55:20.5 & $12.2 \pm 0.38$ & $63.0 \pm 1.96$ \\
IRC133 & 5:41:42.55 & -1:54:03.1 & $10.9 \pm 0.14$ & $56.3 \pm 0.72$ \\
IRC128 & 5:41:45.07 & -1:54:06.5 & $10.5 \pm 0.17$ & $54.1 \pm 0.90$ \\
IRC033 & 5:41:44.41 & -1:55:23.1 & $10.0 \pm 0.43$ & $51.5 \pm 2.23$ \\
IRC058 & 5:41:44.25 & -1:54:55.5 & $8.7 \pm 0.13$ & $45.1 \pm 0.69$ \\
IRC124 & 5:41:42.54 & -1:54:08.8 & $8.2 \pm 0.35$ & $42.5 \pm 1.79$ \\
IRC123 & 5:41:41.64 & -1:54:12.4 & $7.8 \pm 0.14$ & $40.4 \pm 0.71$ \\
IRC158 & 5:41:40.64 & -1:53:48.4 & $6.9 \pm 0.14$ & $35.7 \pm 0.73$ \\
IRC197 & 5:41:43.49 & -1:53:24.7 & $6.8 \pm 0.14$ & $35.3 \pm 0.74$ \\
IRC090 & 5:41:43.25 & -1:54:34.4 & $5.9 \pm 0.14$ & $30.2 \pm 0.70$ \\
IRC180 & 5:41:40.19 & -1:53:34.2 & $5.7 \pm 0.15$ & $29.1 \pm 0.75$ \\
IRC115 & 5:41:42.80 & -1:54:16.5 & $5.5 \pm 0.14$ & $28.2 \pm 0.70$ \\
IRC160 & 5:41:42.61 & -1:53:46.8 & $5.3 \pm 0.14$ & $27.4 \pm 0.73$ \\
IRC059 & 5:41:45.04 & -1:54:54.9 & $5.2 \pm 0.17$ & $26.8 \pm 0.87$ \\
IRC165 & 5:41:41.71 & -1:53:44.4 & $4.6 \pm 0.14$ & $23.5 \pm 0.73$ \\
IRC208 & 5:41:42.34 & -1:53:16.0 & $3.8 \pm 0.15$ & $19.7 \pm 0.78$ \\
IRC175 & 5:41:44.39 & -1:53:36.1 & $3.0 \pm 0.15$ & $15.4 \pm 0.75$ \\
IRC150 & 5:41:43.92 & -1:53:51.5 & $2.9 \pm 0.14$ & $15.0 \pm 0.73$ \\
IRC116 & 5:41:38.23 & -1:54:17.7 & $2.8 \pm 0.14$ & $14.6 \pm 0.72$ \\
IRC144 & 5:41:36.82 & -1:53:59.0 & $2.8 \pm 0.14$ & $14.5 \pm 0.73$ \\
IRC057 & 5:41:45.73 & -1:54:58.1 & $2.7 \pm 0.50$ & $13.8 \pm 2.56$ \\
IRC184 & 5:41:41.35 & -1:53:32.8 & $2.6 \pm 0.14$ & $13.4 \pm 0.74$ \\
IRC120 & 5:41:39.20 & -1:54:14.3 & $2.6 \pm 0.14$ & $13.2 \pm 0.72$ \\
IRC236 & 5:41:44.21 & -1:55:24.0 & $2.5 \pm 0.18$ & $12.9 \pm 0.91$ \\
IRC143 & 5:41:43.56 & -1:53:56.8 & $2.2 \pm 0.14$ & $11.3 \pm 0.72$ \\
IRC206 & 5:41:42.89 & -1:53:16.5 & $2.1 \pm 0.15$ & $11.0 \pm 0.77$ \\
IRC089 & 5:41:43.89 & -1:54:34.2 & $1.9 \pm 0.34$ & $9.8 \pm 1.77$ \\
IRC081 & 5:41:37.75 & -1:54:41.3 & $1.9 \pm 0.13$ & $9.8 \pm 0.69$ \\
IRC103 & 5:41:41.98 & -1:54:24.2 & $1.7 \pm 0.14$ & $9.0 \pm 0.70$ \\
IRC119 & 5:41:36.68 & -1:54:15.3 & $1.3 \pm 0.14$ & $6.7 \pm 0.73$ \\
IRC168 & 5:41:44.12 & -1:53:42.6 & $1.3 \pm 0.14$ & $6.5 \pm 0.74$ \\
IRC170 & 5:41:43.85 & -1:53:38.6 & $1.2 \pm 0.14$ & $6.3 \pm 0.74$ \\
IRC032 & 5:41:44.33 & -1:55:25.0 & $1.2 \pm 0.19$ & $6.2 \pm 0.98$ \\
IRC093 & 5:41:45.63 & -1:54:29.2 & $1.2 \pm 0.36$ & $6.0 \pm 1.86$ \\
IRC080 & 5:41:42.77 & -1:54:39.4 & $1.1 \pm 0.13$ & $5.7 \pm 0.70$ \\
IRC077 & 5:41:43.03 & -1:54:40.6 & $1.1 \pm 0.14$ & $5.5 \pm 0.70$ \\
IRC121 & 5:41:43.46 & -1:54:12.4 & $1.0 \pm 0.14$ & $5.1 \pm 0.71$ \\
IRC131 & 5:41:44.23 & -1:54:05.2 & $1.0 \pm 0.14$ & $5.1 \pm 0.72$ \\
IRC199 & 5:41:36.30 & -1:53:24.7 & $0.9 \pm 0.15$ & $4.8 \pm 0.80$ \\
IRC115B & 5:41:42.80 & -1:54:16.5 & $0.8 \pm 0.14$ & $4.3 \pm 0.71$ \\
IRC025 & 5:41:42.54 & -1:55:37.6 & $0.8 \pm 0.13$ & $4.2 \pm 0.65$ \\
IRC187 & 5:41:43.34 & -1:53:30.2 & $0.8 \pm 0.14$ & $4.1 \pm 0.74$ \\
IRC083 & 5:41:41.49 & -1:54:39.4 & $0.7 \pm 0.13$ & $3.5 \pm 0.69$ \\
IRC074 & 5:41:43.76 & -1:54:41.8 & $0.7 \pm 0.13$ & $3.4 \pm 0.70$ \\
IRC099 & 5:41:44.83 & -1:54:25.3 & $0.6 \pm 0.15$ & $3.3 \pm 0.78$ \\
IRC072 & 5:41:41.39 & -1:54:44.7 & $0.6 \pm 0.13$ & $3.1 \pm 0.69$ \\
IRC194 & 5:41:40.48 & -1:53:27.3 & $0.6 \pm 0.15$ & $3.0 \pm 0.77$ \\
IRC173 & 5:41:36.94 & -1:53:39.5 & $0.6 \pm 0.15$ & $2.9 \pm 0.75$ \\
IRC070 & 5:41:36.83 & -1:54:48.1 & $0.6 \pm 0.13$ & $2.8 \pm 0.69$ \\
IRC045 & 5:41:42.20 & -1:55:10.1 & $0.5 \pm 0.14$ & $2.5 \pm 0.70$ \\
IRC219 & 5:41:43.33 & -1:55:07.0 & $0.5 \pm 0.13$ & $2.5 \pm 0.70$ \\
\hline
\end{tabular}
        \tablefoot{
        \tablefoottext{a}{IRS 2 is contaminated by free-free emission at these wavelengths, and therefore no good estimate of the mass can be made here.}}
\end{table*}

\begin{table}
        \caption{Radii, position angles, and inclinations for resolved disks.}
        \label{tab:resolved}
        \begin{tabular}{llll}
                \hline \hline
                Name & $R_{\text{maj}}$ & Position Angle & Inclination \\
                & AU & deg & deg \\
                \hline
                %IRC101 & 489 $\pm$ 33.5 & 52 $\pm$ 1 & 82 $\pm$ 1.2 \\
                IRC101\tablefootmark{a} & 350 $\pm$ 30 & 45 & 55 \\
                IRS 2 \tablefootmark{b} & 25  $\pm$ 1  & 161 $\pm$ 8 & 36 $\pm$ 4 \\
                        IRC067 & 59 $\pm$ 1 & 144 $\pm$ 12 & 22 $\pm$ 4 \\
                        IRC086 & 48 $\pm$ 2 & 109 $\pm$ 59 & 11 $\pm$ 11 \\
                        IRC044 & 136 $\pm$ 2 & 84 $\pm$ 1 & 70 $\pm$ 1 \\
                        IRC153 & 120 $\pm$ 3 & 89 $\pm$ 1 & 62 $\pm$ 1 \\
                        IRC133 & 50 $\pm$ 2 & 49 $\pm$ 24 & 25 $\pm$ 11 \\
                        \hline
                \end{tabular}
        \tablefoot{
                \tablefoottext{a}{Since IRC101 is not well-described by a Gaussian ellipsoid, we provide a manual estimate of its properties.}
                \tablefoottext{b}{IRS 2 is strongly contaminated by free-free emission}       }
        \end{table}

\subsection{Dust masses of NGC 2024 disks}
\label{sec:dustmassdist}
\begin{figure}[ht]
\begin{center}
                \includegraphics[width=0.49\textwidth]{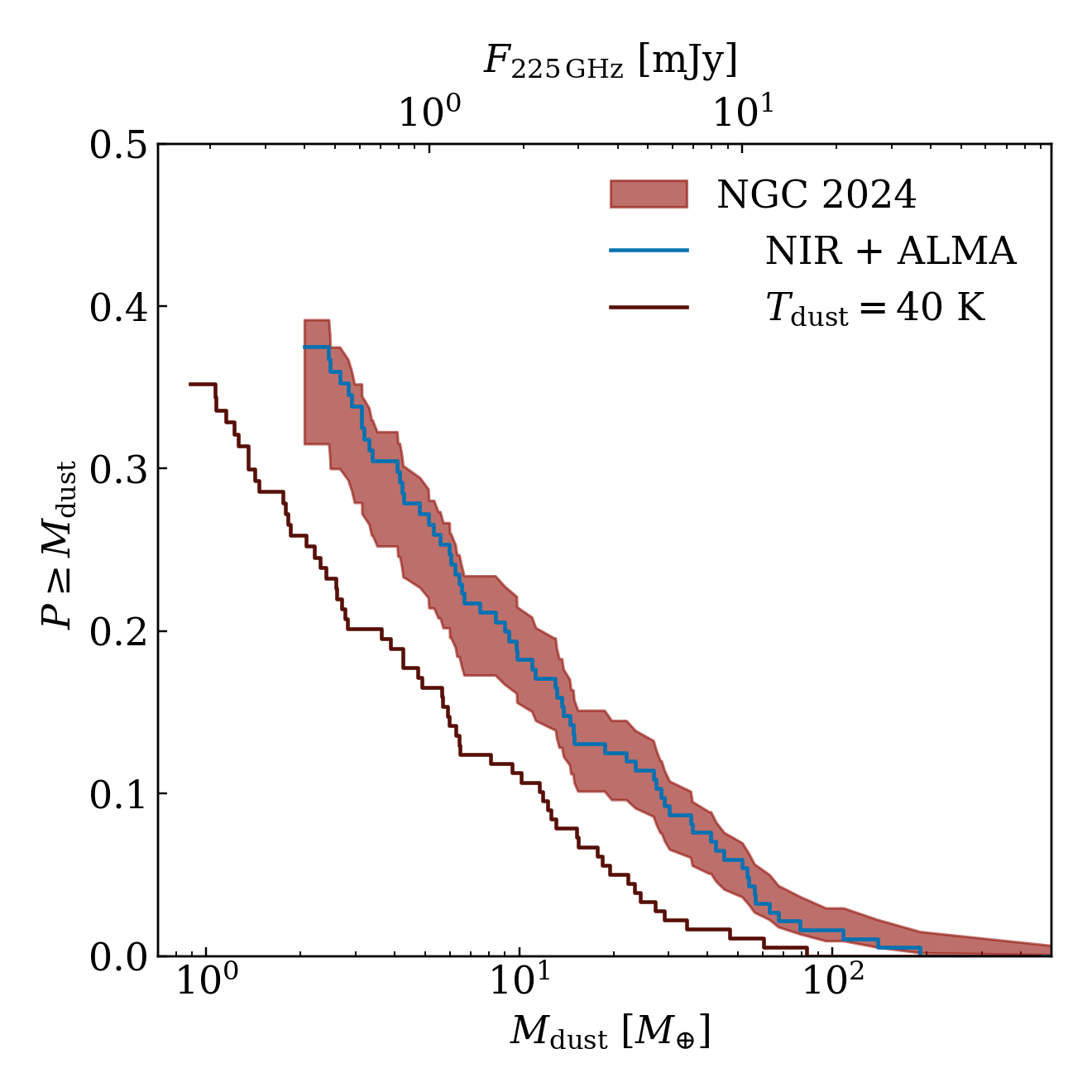}
                \caption{Inferred disk mass distribution for the NGC 2024 disk population (dark red). The effect of adding sources observed with ALMA but not detected in previous observations of NGC 2024, overlapping the~\citet{meyer96} catalog, and consistent with being Class II sources is shown in blue. The dark red line shows the effect of assuming a 40\,K temperature for the sample.}
                \label{fig:mdisk_v1}
\end{center}
\end{figure}

If the continuum emission detected in the sample of protoplanetary disks studied here is optically thin, a simple relation exists between the mass (of millimeter-sized dust grains) of the disk and its flux:
\begin{equation}
M_{\text{dust}} = \frac{d^2 F_{\nu,\,\text{dust}}}{\kappa_{\nu} B_{\nu}(T_{\text{eff}})}.
\label{eq:thindisk}
\end{equation}
In Fig.~\ref{fig:mdisk_v1}, the Kaplan-Meier estimator is used to infer the disk mass distribution for the full sample, including nondetections. In this article, we take the distance $d$ to NGC 2024 to be 414\,pc~\citep{menten07, bailerjones18}. To facilitate the comparison of this sample of disks to other ALMA surveys of disks in star-forming regions~\citep[e.g.,][]{ansdell16,vanterwisga19}, we use standard assumptions for the values of the other parameters in Eq.~\ref{eq:thindisk}: $T_{\text{eff}} = 20$\,K~\citep{andrews05}, and $\kappa_{\nu} = \kappa_0 \left( \nu / \nu_0 \right)^{\beta}$ with $\beta = 1$ and $\kappa_{1000\,\text{GHz}} = 10$\,cm$^{2}$\,g$^{-1}$~\citep{beckwith90}.

The assumption that all disks are optically thin in Band 6 is currently a topic of active debate. Several results have indicated that disks may be partly optically thick~\citep{tripathi17,andrews18,zhu19}, while it has also been suggested that the method used here can lead to the dust mass being overestimated~\citep{rosotti19}. These effects are difficult to quantify even for well-studied disks. Since the primary purpose of these observations is to study how the continuum luminosity of a disk changes as the solids evolve over time, we therefore also show a flux axis in the disk mass distribution plots shown here, which is free from assumptions on opacities and optical depth.

Our use of the Kaplan-Meier estimator requires that the inclusion of an object in the catalog is not sensitive to its disk mass, that the sample is drawn from a single population, and that the probability of a nondetection does not depend on the variable studied (here, the disk mass). While the first of these criteria is met, the second does not necessarily hold, and we examine it further in Sect.~\ref{sec:internal}. The third requirement is not met, and means these results should be interpreted carefully at the lower end of the mass range, where completeness begins to drop~\citep{mohanty13}. 

\subsubsection{Impact of free-free emission}
By using Eq.~\ref{eq:thindisk}, we implicitly assume no free-free emission contaminates our observations. In strongly irradiated star-forming environments, however, this assumption does not necessarily hold. Proplyds in the Trapezium cluster, for instance, have non-negligible contamination from free-free emission even in Band 6 based on observations at centimeter wavelengths~\citep{mann14,eisner18}. However, such concerns do not seem to be significant in NGC 2024. For example,~\citet{mann15} found free-free emission contributes only weakly to the flux of the majority of sources at $338.2$\,GHz. The contribution from free-free emission is $10\%$ for the most contaminated object, IRC 065, which is located out of the field covered here, and $<1\%$ for the most contaminated sources that are also covered by this survey (IRC 101 and IRC 071). For these sources, any free-free emission should still also be negligible in our observations, assuming spectral indices of $2.3$ for the dust and $-0.1$ for the free-free emission (based on~\citealp{ricci10a,ricci10b,tychoniec18}). While VLA data only exist for part of the field we observed with ALMA -- roughly, the southeastern quarter of the area covered here -- this field is the part of the image with sources closest to IRS-2b. 
\subsubsection{Impact of midplane temperature variations}
The use of a 20\,K effective temperature for the disk midplane is also worth considering, even apart from the issue of optical depths. This value is an average effective midplane temperature originally found by fitting SEDs with disk models with 100\,AU radii~\citep{andrews05}. As a result it may not be appropriate in regions where disks are compact due to external photoevaporation. Indeed, such truncated disks tend to be better described by higher effective temperatures, as shown in~\citet{eisner18}. Additionally, the ambient radiation field may have an impact on the disk temperature. These effects are difficult to constrain in our observations alone, but would both lead to an overestimate of the dust mass of each disk. The size of this effect is shown in Fig.~\ref{fig:mdisk_v1} for a 40\,K effective midplane temperature, which is found to be an appropriate value for low-mass compact disks with a characteristic radius $R_{\text{c}}$ of 25\,AU in the grid of models in~\citet{eisner18}.

\subsubsection{Point sources without near infrared counterparts}
\label{sec:newsources}
Twelve sources are clearly detected ($>4.8\,\sigma$) but do not correspond to Class-II YSOs identified previously. These sources were first identified by performing a blind search for $>4.8\,\sigma$ peaks in the data, after which their presence in both the long-baseline image and the full-baseline data was checked. Finally, we performed a visual inspection to confirm the reality of these sources. We present their positions and fluxes in Table~\ref{tab:U}. Cutouts of these sources can be found in Appendix~\ref{app:cutouts}, in Fig.~\ref{fig:cutouts3}. The significance cut used here was chosen such that we expect less than one false positive given the number of independent beams in the full image (0.6 on average). These otherwise unidentified sources are marked in the overview map in Fig.~\ref{fig:disksinspace} with open plus signs. Of these sources, ten overlap with the area covered by the~\citet{meyer96} United Kingdom Infra-Red Telescope survey. The remaining two sources (U12 and U13) are located in the northeast of the map, just above the dense molecular ridge hosting the FIR 1 - 4 sources.

Without a more complete SED, determining the nature of these additional sources is difficult. Only one source, U13, corresponds to a member of the Massive Young Star-Forming Complex Study in Infrared and X-ray (MYSTiX) catalog~\citep{povich13}. Several of the other sources not in the catalog show interesting properties and are discussed in more detail (along with the extraordinary IRC 101 transition disk) in Sect.~\ref{sec:individualobjects}. Some are clearly not Class-II sources based on their morphology (for example, U2 and U3) or because they coincide with a known FIR source (U5 is associated with the known FIR-3 source). We have conservatively assumed that the rest of these objects are disks. By including those disks that do fall within the original~\citet{meyer96} field in the sample, the detection rate is biased, but in an informative way: as it is unknown how many sources are missing in the catalog that are nondetections in ALMA, this gives the highest possible value for the `true' detection rate of disks in this area, of $35 \pm 4\%$. As Fig.~\ref{fig:mdisk_v1} shows, this effect is small. The low number of new detections also implies that there is no large population of disks more massive than 3.6\,$M_{\oplus}$ in the sample.

\begin{table}
\caption{Continuum fluxes for objects not included in the base catalog.}
\label{tab:U}
\centering
\begin{tabular}{l l l l l}
\hline \hline
Name & RA & Dec & Flux & D \\
 &  &  & mJy &  \\
\hline
U1 & 0:22:47.02 & -1:55:31.6 & $10.3 \pm 0.80$ & Y \\
U2 & 0:22:47.01 & -1:55:30.9 & $100.5 \pm 1.65$ & N \\
U3 & 0:22:46.71 & -1:55:07.3 & $8.4 \pm 0.40$ & N \\
U4 & 0:22:46.51 & -1:54:37.0 & $0.6 \pm 0.13$ & Y \\
U5 & 0:22:46.87 & -1:54:26.4 & $242.1 \pm 0.41$ & N \\
U6 & 0:22:46.88 & -1:54:25.4 & $11.0 \pm 0.34$ & Y \\
U7 & 0:22:46.71 & -1:54:07.6 & $1.4 \pm 0.13$ & Y \\
U8 & 0:22:46.86 & -1:53:43.5 & $1.6 \pm 0.14$ & Y \\
U9 & 0:22:46.77 & -1:53:38.2 & $1.8 \pm 0.14$ & Y \\
U10 & 0:22:46.76 & -1:53:32.8 & $3.8 \pm 0.14$ & N \\
U11\tablefootmark{a} & 0:22:46.82 & -1:53:07.6 & $41.3 \pm 0.62$ & Y \\
U12\tablefootmark{a} & 0:22:46.93 & -1:53:02.3 & $1.7 \pm 0.33$ & Y \\
\hline
\end{tabular}
\tablefoot{
\tablefoottext{a}{Source not in the area covered by~\citet{meyer96}.}
}
\end{table}

\subsubsection{Stacking analysis of nondetections}
\begin{figure}[ht]
\begin{center}
                \includegraphics[width=0.49\textwidth]{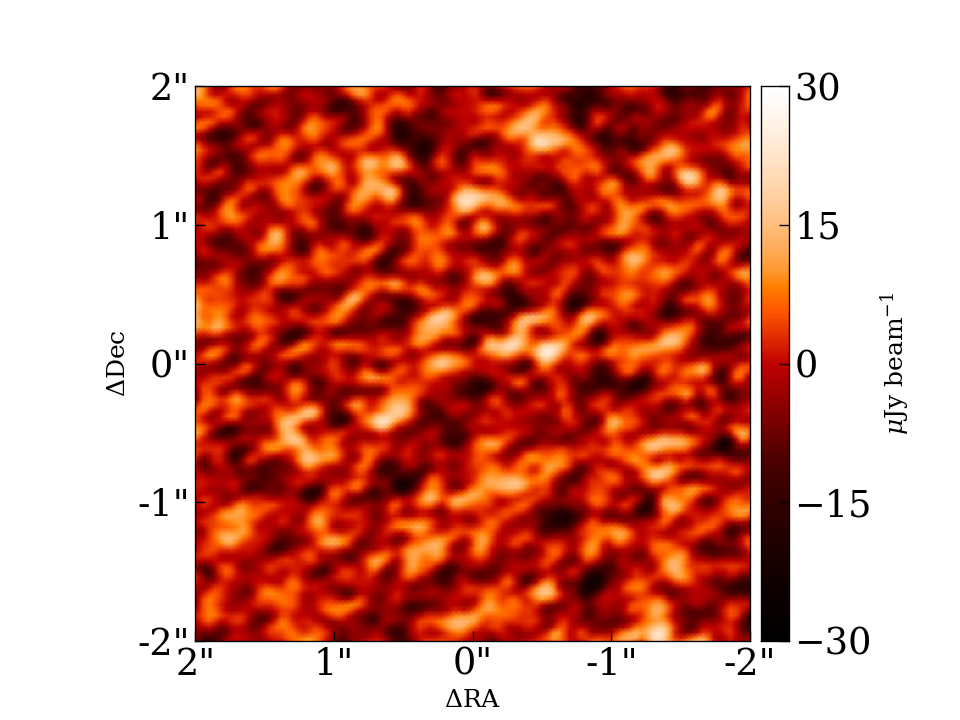}
                \caption{Noise-weighted stacked image of nondetections in the ALMA observations.}
                \label{fig:stack}
\end{center}
\end{figure}

In total, 122 catalog objects across the image were not detected. By stacking all these sources, in theory, an SNR improvement of more than a factor of ten can be achieved.
The stacking analysis was performed by masking out the apertures containing detected sources in the full image, and taking the average (weighted by the local noise) of all positions not associated with a millimeter continuum source. The resulting image, shown in Fig.~\ref{fig:stack} has a noise level of $6.66\,\mu$Jy\,beam$^{-1}$. No source is detected in this stacked image. The choice of masking out detected sources is necessary because several sources (like IRC 101 and FIR 3) are so bright that their emission contaminates the final image, but does not significantly effect the derived mass limit.

The resulting $3\sigma$ mass limit for the stacked sources is $20\,\mu$Jy, or $< 0.96\,M_{\text{Mars}}$. This implies that on average these disks will not be able to form even a single Mars-sized planet embryo from millimeter-sized grains currently present in the disk. This analysis, however, does not exclude the possibility that such bodies have already formed. This low upper limit is consistent with the disk mass distribution inferred in the region.

\subsection{Comments on individual objects}
\label{sec:individualobjects}
Several objects in the observed field are of particular interest. Here, we discuss the continuum properties of the largest resolved disk in the sample. Several younger (Class 0) sources are also present in this sample and observed at higher resolutions than previously available. Finally, two compact sources of nonthermal emission are identified.

\begin{figure}[ht]
\centering
                \includegraphics[width=0.49\textwidth]{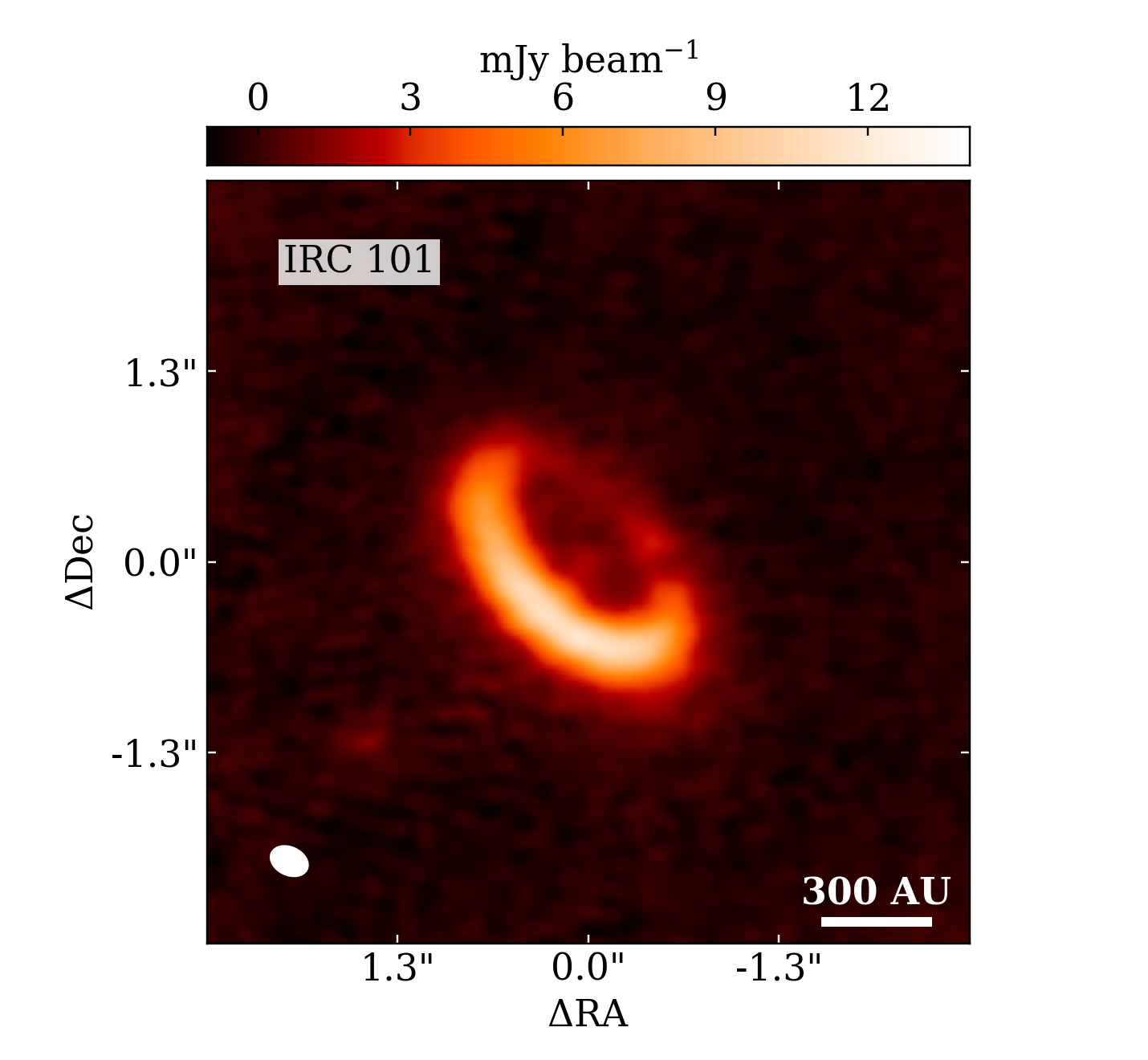}
                \caption{Zoomed-in panel from the full-baseline ALMA map of NGC 2024 showing an apparently asymmetric dust ring around IRC 101. It is resolved for the first time in these observations.}
                \label{fig:onepanel}
\end{figure}
\begin{figure}[ht]
\centering
                \includegraphics[width=0.49\textwidth]{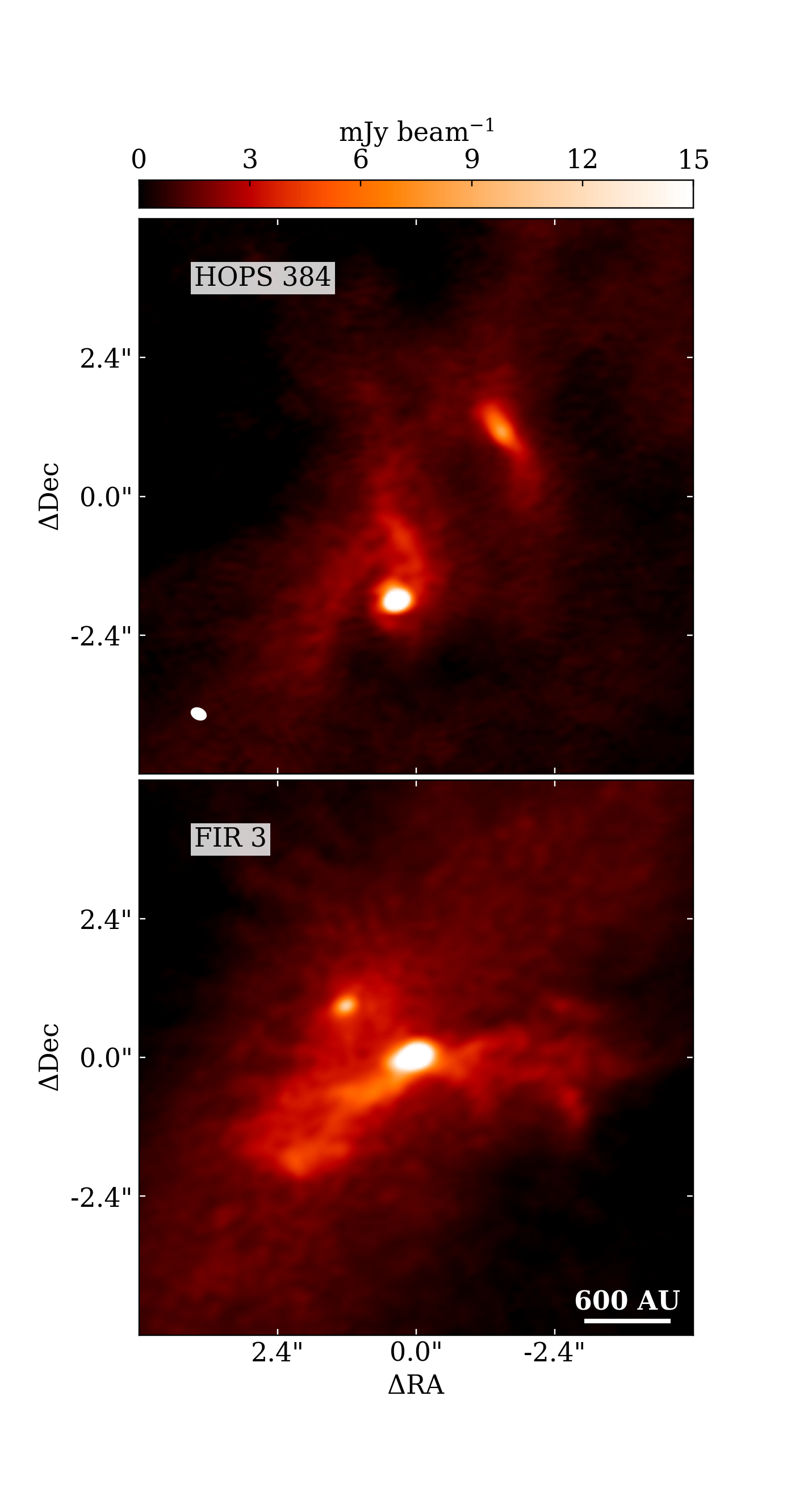}
                \caption{Zoomed-in panels from the full-baseline ALMA map of NGC 2024 showing two resolved, embedded objects with particularly interesting features. HOPS 384 (top) is a Class 0 protostar, here resolved as two, possibly interacting, YSOs. FIR 3 (bottom) is resolved for the first time in these observations, and shows two outflow cavity walls in 225\,GHz continuum observations, as well as two compact continuum sources.}
                \label{fig:twopanel}
\end{figure}

\begin{figure}[ht]
\centering
        \includegraphics[width=0.49\textwidth]{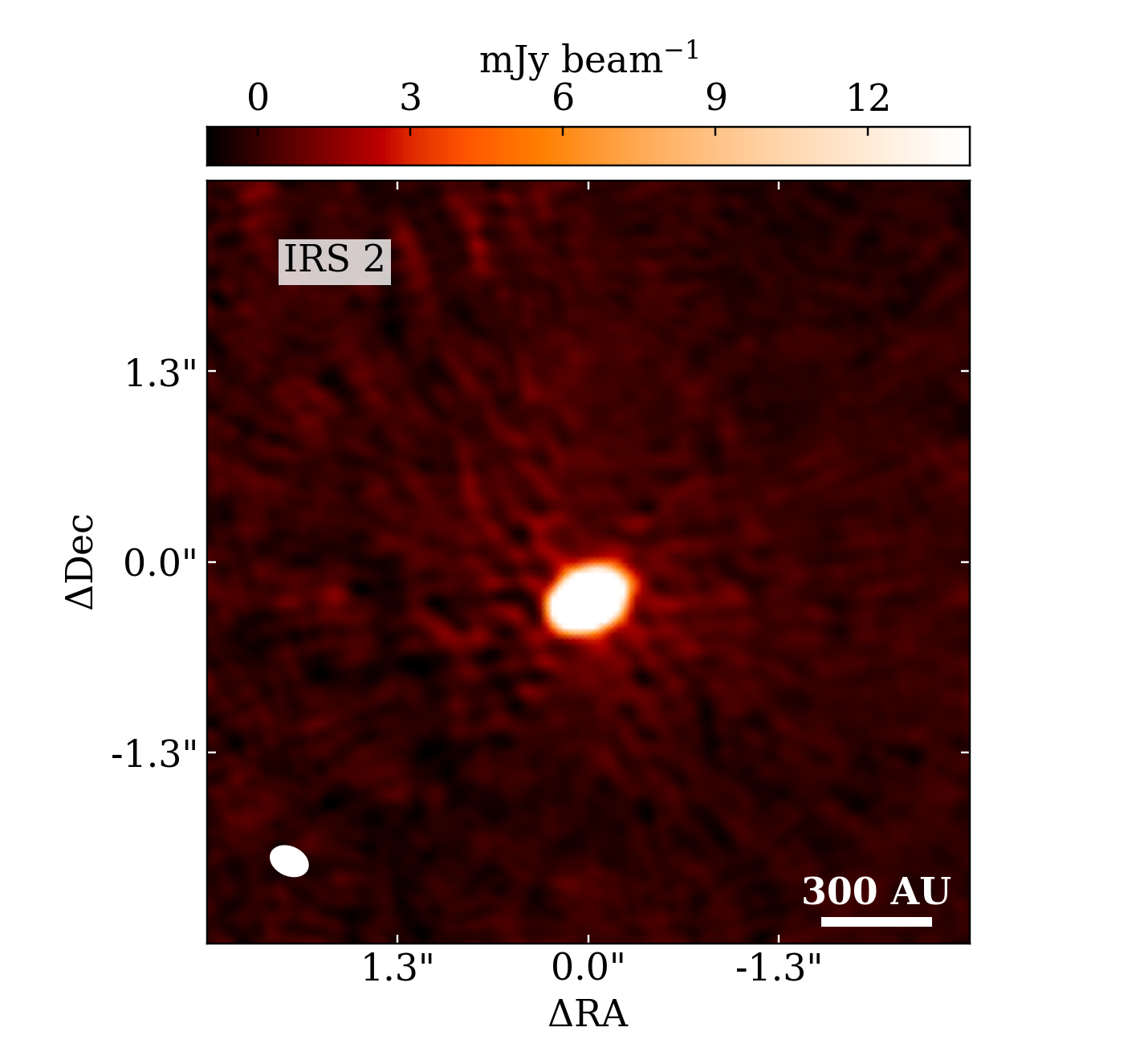}
        \caption{Zoomed-in panel from the full-baseline ALMA map of NGC 2024 showing the resolved disk of the massive IRS 2 protostar.}
        \label{fig:IRS2zoom}
\end{figure}
\subsubsection{IRC 101: A 300\,AU-radius dust ring around a YSO}
\citet{eisner03} and~\citet{mann15} already identified this object as an extraordinarily bright YSO at millimeter wavelengths, and the latter group suggested that since the peak flux did not match the infrared catalog positions precisely, its emission might originate from the envelope around a Class 0 protostar. In Fig.~\ref{fig:onepanel}, the first resolved ALMA observations of this source are shown. Thanks to the excellent resolution of the data, the source is revealed to have a morphology similar to that of `classical' transition disks, with a well-defined inner cavity. Its size, however, remains extraordinary: assuming a 414\,pc distance to NGC 2024 implies that the ring's peak intensity is at a radius of $\sim 300$\,AU, making it the largest such object identified so far. An inner disk may also be present, but the central emission is faint and unresolved. Such inner disks are also seen in some other transition disks (see, for instance, the sample in \citealp{francis20}). Given the size and depth of the cavity, which is well resolved in these observations, IRC 101's disk is likely to host a multiple star system, rather than being carved by a planetary-mass companion. Additional evidence for this scenario comes from the apparent asymmetry in the ring, which is significantly brighter in the south than in the north. GG Tau A has a circumternary dust ring with a peak radius of 229\,AU, and shows strikingly similar asymmetric continuum emission~\citep{tang16}.

Unfortunately, the SED of this fascinating object is not well sampled, due to the high optical extinction and its proximity on the sky to IRS 2b. The disk is, however, detected in several channels in $^{13}$CO confirming that its systemic velocity is consistent with membership of NGC 2024 and excluding the possibility that this is a more compact foreground object.

\subsubsection{NGC 2024 FIR 3}
These observations also resolve for the first time a deeply embedded YSO, coincident with the location of NGC 2024 FIR 3, as well as a second continuum source nearby. Both of these sources are shown in Fig.~\ref{fig:twopanel}, in the bottom panel. Clearly, these sources (which lack counterparts in NIR observations) are young. The brightest source shows two approximately symmetric continuum structures, which we here interpret as outflow cavity walls, extending over more than 1000\,AU to either side and showing irregular, clumpy substructures. Interestingly, the continuum emission around this YSO seems to be somewhat resolved, and is slightly elongated to the west. Previously,~\citet{ren16} have suggested NGC 2024 FIR 3 may be a first hydrostatic core. Our observations are inconsistent with that hypothesis, given the resolved nature of the outflow cavity walls and central source~\citep{young19}.

\subsubsection{HOPS-384: A very young multiple system?}
HOPS-384, shown in the top panel of Fig.~\ref{fig:twopanel}, is a known Class 0 protostar~\citep{furlan16}. In these observations, we resolve it as two objects, both associated with significant extended continuum emission. The brightest source is in the south, and seems to be associated with two asymmetric, spiral-arm-like curved arcs of dust. The fainter northern component has an elongated, s-like shape, and seems to coincide with an extended ridge of continuum emission, although the lack of short baseline coverage in these observations prevents us from making this association more explicit. The northern source seems to be associated with more compact structures than the southern source. It is not clear if a physical link between them exists; certainly, the arcs seen in both sources are curved in the same direction.

\subsubsection{IRS 2}
IRS 2 is, like IRC 101, a prominent source at millimeter wavelengths, with a continuum flux of $241 \pm 32$\,mJy. Earlier work has shown that IRS 2 is a massive young star of $\sim 15\,M_{\odot}$ and that it is surrounded by an inner gaseous disk and a dust disk~\citep{lenorzer04, carattiogaratti20}. The disk is resolved in our ALMA observations; a cutout is presented in Fig.~\ref{fig:IRS2zoom}. Interestingly, the inclination and position angle of the disk that we infer using the \textit{imfit} task in \texttt{CASA} are indistinguishable to within their respective errors from the values found by~\citet{carattiogaratti20} using GRAVITY/VLTI-observations of the inner disk regions. This suggests that the inner and outer components of the disk are located in the same plane.

Given the high mass of the central star, we tested if there was evidence of nonthermal emission. The main limitation to calculating an in-band spectral index with these data is that we have only limited spectral leverage between the lowest and highest frequency continuum spectral windows, which causes the built-in Taylor-expansion algorithm in the \textit{tclean} task to fail. A second issue is that the relative flux calibration between the spectral windows introduces an unknown absolute error. Finally, the position of IRS 2 at the edge of the mosaic means the local noise level is rather high.

We imaged the pointings surrounding IRS 2 including the bright IRC 101 disk in the lowest  and highest frequency continuum spectral windows, and manually calculated the spectral indices of both objects. We find that IRS 2 has a significantly lower spectral index than IRC 101, which we infer to be the result of free-free contamination of the emission in IRS 2. To be cautious, we do not include the mass of this source in the subsequent analysis for this reason.

\subsubsection{Bright flaring object at 225\,GHz}
\label{sec:flare}
\begin{figure*}[ht]
\begin{center}
                \includegraphics[width=\textwidth]{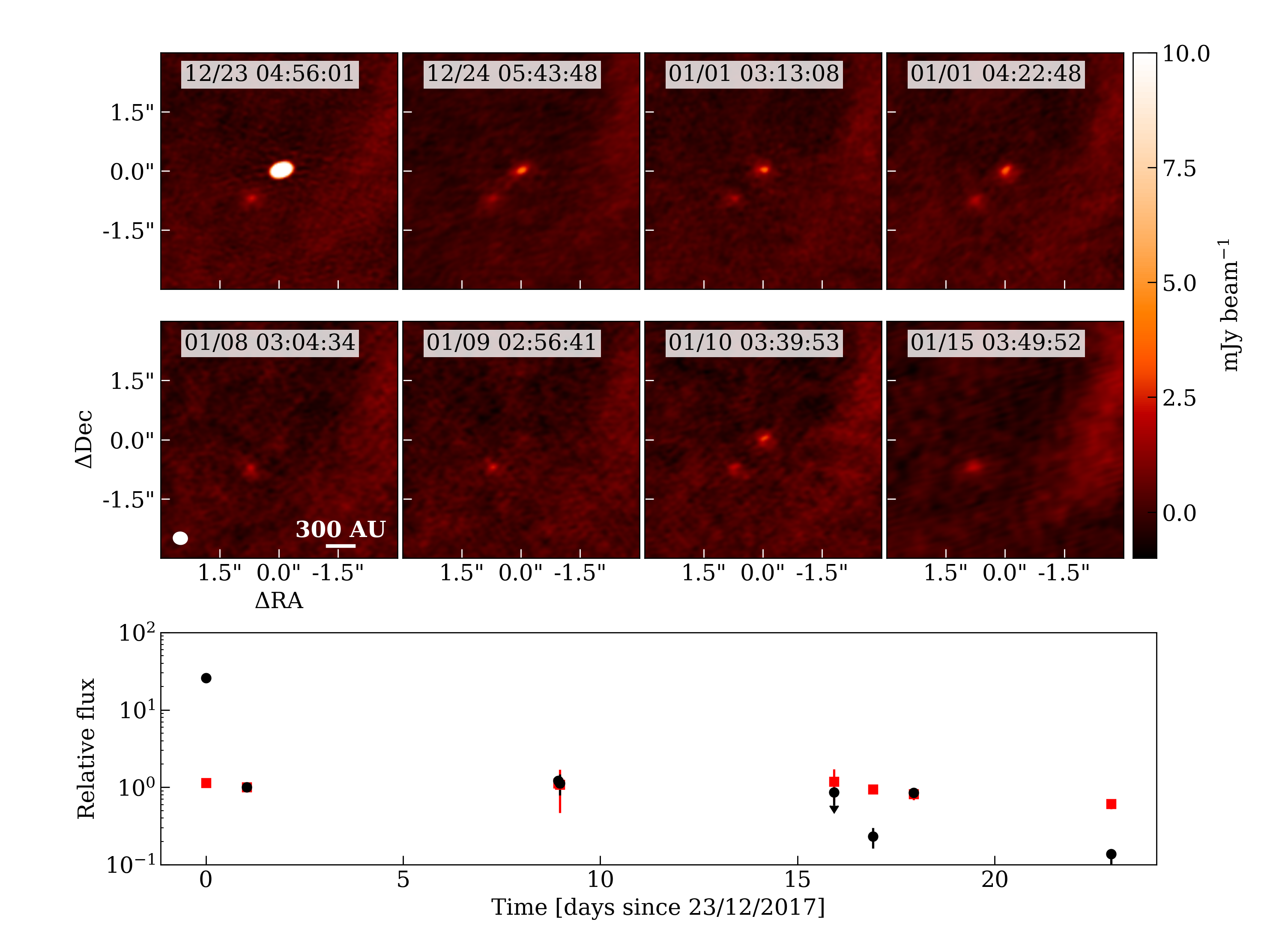}
                \caption{Time-resolved observations of a flaring YSO. Top: Zoomed-in panels of the U11 source from individual integrations of the ALMA continuum show the source variability relative to IRC 184. Bottom: Fluxes of both IRC 184 (red) and U11 (black), normalized to these objects' fluxes on the second full integration of the mosaic.}
                \label{fig:flarevsnoflare}
\end{center}
\end{figure*}

Given the observation schedule used, the full ALMA field was observed on multiple nights, allowing us to test if any objects showed signs of variability. Indeed, J054141.3-015332, U10 in this catalog, shows signs of extreme variability on short ($\leq 24$-hour) timescales. It is the brightest continuum source in the field on the first day of observing with a total flux of $176.9 \pm 0.5$\,mJy. It then fades by a factor of 26 to $6.8 \pm 0.36$\,mJy within 24 hours, and shows lower amplitude flickering in the subsequent integrations. The source flux falls below the detection limit in the fifth and eighth full integrations, as shown in Table~\ref{tab:flare}. Figure~\ref{fig:flarevsnoflare} shows the flux of the flaring source and the adjacent IRC 184, normalized to the second full integration.%
\begin{table}
\caption{Flux over time for J054141.3-015332, starting December 23, 2017.}
\label{tab:flare}
\centering
\begin{tabular}{l l l}
\hline \hline
Day & Time & F$_{225\,\text{GHz}}$ \\
    &      & mJy \\
\hline
12/23 & 04:56:01 & $176.9 \pm 0.5$ \\
12/24 & 05:43:48 & $6.8 \pm 0.4$ \\
01/01 & 03:13:08 & $8.3 \pm 0.4$ \\
01/01 & 04:22:48 & $7.7 \pm 2.31$ \\
01/08 & 03:04:34 & $< 5.9$ \\
01/09 & 02:56:41 & $1.6 \pm 0.5$ \\
01/10 & 03:39:53 & $5.8 \pm 0.5$ \\
01/15 & 03:49:52 & $< 0.9$ \\
\hline
\end{tabular}
\end{table}
The timescales on which the source shows variability are inconsistent with the variability being dominated by blackbody emission, since they require emission from a very compact area. The finding of millimeter-variability for this source agrees with earlier observations. For example, in X-rays this source is among the brightest X-ray emitters in the region and may be variable~\citep{skinner03}. The SED of the emitting source, however, is uncertain: due to the close proximity on the sky to IRC 184, it is not clear if the infrared excess is present in both sources, or in only IRC 184 or U11.

\section{Discussion}
The primary goal of these observations was to improve our understanding of how disk properties vary across NGC 2024, and how they compare to those in other star-forming regions of different ages and different (F)UV fields. In Sect.~\ref{sec:internal}, we therefore define two subsamples and interpret their disk mass distributions in terms of the physical properties of NGC 2024.

%[First NGC 2024 internally, then: comparison to other regions]
\subsection{Two disk populations across the NGC 2024 core region}
\label{sec:internal}
The disk mass distribution in Fig.~\ref{fig:mdisk_v1} implicitly assumes a single population of disks is observed in this field. This assumption is not necessarily true, however, given the complex environment covered in these observations. The radiation field of IRS 2b is important throughout the region, even in the vicinity of IRS 1~\citep{burgh12}, although the presence of large amounts of foreground extinction makes it less easy to observe than the Trapezium. Projected distances of YSOs in the catalog to IRS 2b range from 0.0057\,pc to 0.36\,pc. In the Trapezium cluster~\citep{odell01,abel19}, disk masses vary strongly across this distance range due to external photoionization. There, disks nearest to the ionizing source ($\theta^1$\,Ori\,C) have significantly lower masses~\citep{mann14,eisner18}. A second reason that disk properties may vary across the ALMA image is an unequal distribution of interstellar material, which may attenuate ionizing radiation from the massive young stars (if sufficiently dense) or host very young populations. Indeed, the area covered in this survey is dominated on the eastern side by the dense molecular ridge hosting the FIR sources~\citep{watanabe08}, while the western side seems to lack dense interstellar material. Finally, there is evidence of a core-halo age gradient in NGC 2024, with the youngest stars located in the cluster's core, but a rapid increase in age (from 0.2\,Myr to 1.1\,Myr) in the inner 0.5\,pc~\citep{haisch00, getman14}. This age estimate is based on the J-band photospheric emission and the X-ray emission, and should therefore be independent of the millimeter flux properties. All these factors may influence disk masses over space within the observations presented here.

In Fig.~\ref{fig:disksinspace}, where detections and nondetections are shown against both the ALMA data and Herschel PACS 160\,$\mu$m contours~\citep{stutz13}, the hypothesis that two distinct populations of disks are indeed present in this survey immediately suggests itself: targets located in the east, along the direction of the dense molecular ridge, seem to have a much higher detection rate than those in the west. This eastern region also contains all known Class I and Class 0 sources in the field. The dividing line between these populations seems to lie on a mildly inclined north-south axis.
\begin{figure}[ht]
\begin{center}
                \includegraphics[width=0.49\textwidth]{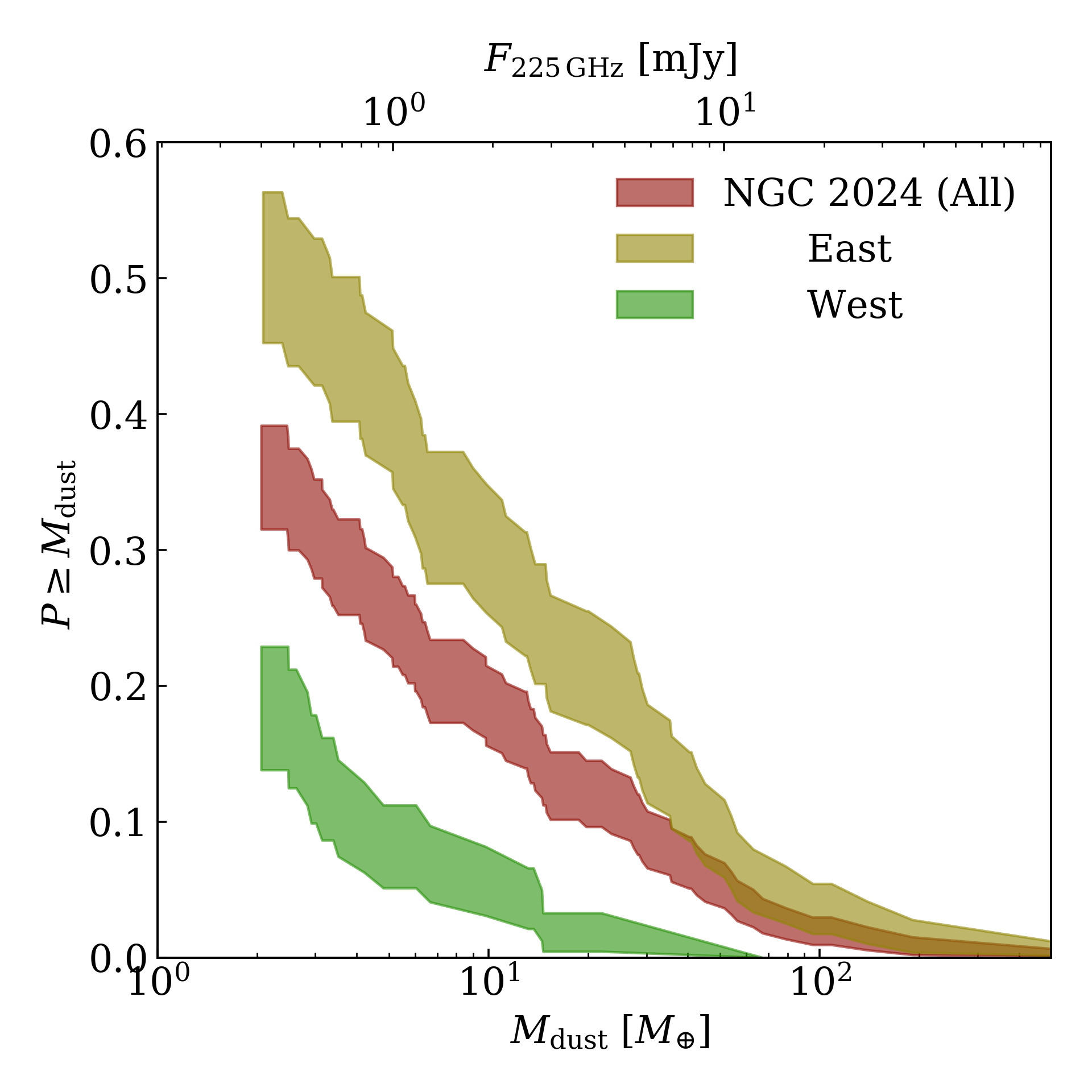}
                \caption{Inferred disk mass distribution for the NGC 2024 disk population (dark red) and the hypothesized western and eastern subpopulations, in tan and green respectively. The dividing line between these populations runs from 5:41:44.189 -1$^{\circ}$55'39.94'' and 5:41:38.665 -1$^{\circ}$53'00.27'', tangentially to the southern part of the dense molecular ridge.}
                \label{fig:twopop}
\end{center}
\end{figure}
To test this hypothesis formally, we divided the sample into two subsamples in a simple way. For both subsamples, the disk mass distribution was calculated separately in the manner described in Sect.~\ref{sec:dustmassdist}. Using a log-rank test, we then tested if it is possible to reject the null hypothesis that there is no difference in disk masses between the samples. The subsamples are divided by a line between 5:41:44.189 -1$^{\circ}$55'39.94'' and 5:41:38.665 -1$^{\circ}$53'00.27'' (J2000). The eastern subsample (which we will refer to as NGC 2024 East in the following) has $N = 97$ while the western (NGC 2024 West) has $N = 82$. In observational terms, the dividing line is drawn along the 4\,Jy\,pixel$^{-1}$ contour in the 160\,$\mu$m data, just to the south of the densest part of the ridge. The physical motivation for this division is that it should be tangent to the densest part of the molecular ridge containing the FIR 1--5 sources, and ensure all those sources are in one subsample. While this division is to a certain extent arbitrary, we have tested that the final results are not sensitive to its position, and that they hold so long as the dividing line lies within $10^{\circ}$ of its presently defined angle, and less than $0.5'$ to the west of its current position.

In Fig.~\ref{fig:twopop}, the disk mass distributions of the resulting subsamples are shown. The log-rank test indicates that these distributions are extremely unlikely to be drawn from the same population ($p = 2.0 \times 10^{-6}$). This result is striking: the eastern population has the smallest projected distances to the location of the ionizing source, but has significantly higher masses, and a detection rate of $45 \pm 7\%$. In contrast, only $15 \pm 4\%$ of disks in the western part of the image are detected. This result is at odds with a view where only external photoevaporation determines disk mass in this region.~\citet{mann15}, who observed the eastern disk population with the Submillimeter Array, similarly did not see evidence of external photoevaporation in that part of the nebula. It is therefore important to interpret this result in the context of previous observations of NGC 2024, to arrive at a more detailed view of the structure and history of this star-forming region.

\begin{figure}[ht]
\begin{center}
                \includegraphics[width=0.49\textwidth]{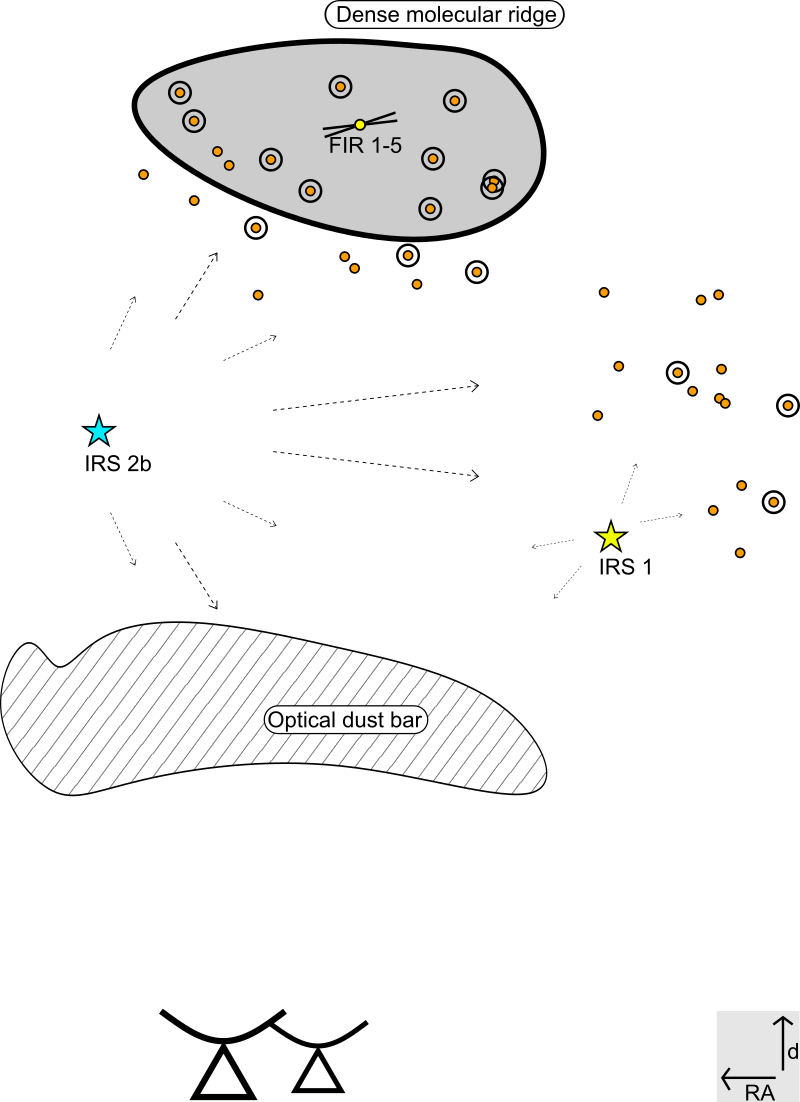}
                \caption{Schematic view of the young stellar populations in NGC 2024. Orange points are Class II stars; circles indicate millimeter detections. The eastern population (left) is young and located near the dense molecular ridge hosting the FIR 1--5 Class 0 sources, which partly shields it from the intense radiation field driven by the ionizing source(s), such as IRS 2b. The western population is older, and more exposed to the ionizing radiation of the primary ionizing source(s) and the cooler IRS 1.}
                \label{fig:schematic}
\end{center}
\end{figure}

Here, we propose that two effects contribute to the observed properties of the entire sample. To the east, we look toward a very young disk population, still mostly embedded in a dense molecular ridge, and quite far (in projected distance) from IRS 2b. To the west, the observed disks are not only older, but much more exposed to radiation from the ionizing source(s) and IRS 1, resulting in lower disk masses. This scenario is illustrated schematically in Fig.~\ref{fig:schematic}. This view builds on previous studies of NGC 2024's molecular environment using a blister model of expanding HII regions, which provide the information on how the various cloud components are ordered along the line of sight. In the current view, an optical dust bar is placed in front of the main ionizing sources along the line of sight. The massive stars in the cluster core excavate an expanding HII blister, impacting a dense ridge of cold molecular gas behind it~\citep{giannini00, emprechtinger09}. At this point, unfortunately, no detailed information on the line-of-sight distances toward the different components is available, not even from Gaia Data Release 2.

Our proposed young, eastern population would be located in and shielded by the dense ridge. Indeed, several YSOs in this region show clear signs of youth. For example, FIR 3 has no NIR counterpart and visible outflow walls in millimeter-continuum observations. Also, HOPS-384 has been found to be a Class 0 source from previous SED fitting~\citep{furlan16}. The ages of the stars with disks are also low in the cluster's core~\citep{getman14}. We note that in the integral-shaped filament in Orion A, the youngest sources are found to be clustered along the filament, while stars with disks are not~\citep{kainulainen17}. The integral shaped filament is likely to be somewhat older than NGC 2024, and the dispersal of Class II disks that is seen there takes time. We therefore suggest that the objects in the eastern part of our map of NGC 2024 are young and still at least somewhat shielded by the dense molecular ridge in which they formed. This picture is also in agreement with the higher extinction values toward this region~\citep{lombardi14}.

In the western population, the extinction along the line of sight is much lower. This population has both higher stellar ages~\citep{getman14} and a lower disk fraction~\citep{haisch00}, which is in line with the expectation for a sample of stars that have been exposed to intense ionizing radiation fields for $\sim 1$\,Myr. These observations reveal that disk masses in this part of the cluster are lower, too. It is interesting to note that while one or more stars near the position of IRS 2b still dominate the total FUV radiation budget in the nebula, the somewhat cooler B0.5 star IRS 1 contributes a significant amount of flux even at $1200\,\AA$~\citep{burgh12}. This other star may help increase the efficiency of external photoevaporation, if it is driven mainly by FUV irradiation of the disks in this part of the nebula. In Fig.~\ref{fig:compop}, the disk mass distributions of the two disk populations in NGC 2024 are compared to the disk mass distributions in isolated star-forming regions of various ages, and to disks in regions where we know external photoevaporation is significant. As the figure shows, there is a degeneracy between the effects of region age and photoevaporation. Thus, the low disk masses in this region could also be explained by age alone. Despite the large uncertainties inherent in YSO age estimates, the western population's disk masses most closely resemble the $6-10$\,Myr-old disks of Upper Sco~\citep{barenfeld16} (see Sect.~\ref{sec:westvelsewhere}), making it more likely that it is age together with external photoevaporation that causes the low observed disk masses.

The stark difference in disk masses between the two populations identified here indicates a complex star formation history. In particular, it is possible that the younger population is the result of a moving region of star formation as the HII bubble is carved out and compresses the gas behind it. The western population might then be the product of an earlier period of star formation. To test this prediction, deep spectroscopic observations of the YSOs in the eastern sample, or covering a wider sample of NGC 2024 disks with ALMA, are necessary.

\subsubsection{Impact of uncertain stellar masses}
\label{sec:mstar}
The interpretation of the different disk masses in the eastern and western populations identified here is based on the assumption that there are no significant differences in the masses and multiplicities of the stars observed here. The higher extinction to the east, however, would make it more difficult to detect lower mass stars, in effect artificially biasing the stellar population to a higher average mass in the east. This is important: a clear correlation between disk mass (or luminosity) and stellar mass of the form $L_{\text{mm}} \propto M_{\star}^{1.3-2.0}$ has been demonstrated in many nearby star-forming regions~\citep[e.g.,][]{andrews13,pascucci16,ansdell17}. This relation, however, may not hold in a region with significant photoevaporation. If NGC 2024 as a whole is like the ONC, we might not expect a strong dependence of $M_{\text{disk}}$ with $M_{\star}$, or none at all~\citep{eisner18}. At this point it is not clear how well this result generalizes to other regions such as NGC 2024.

The~\citet{meyer96} catalog used to define this sample is very deep: it samples stellar masses down to $0.1 M_{\oplus}$ viewed through $A_V = 19$. This depth helps minimize the problem of nonuniform stellar masses, but not completely.
More importantly, the scatter in $M_{\text{disk}}$ versus $M_{\oplus}$ is generally very large, more than 1.5\,dex, in nearby SFRs~\citep{pascucci16}, which implies that we should detect at least some of the disks around even the lowest mass stars. As discussed in Sect.~\ref{sec:newsources}, several objects that are consistent with being disks around stars missing from the catalog are in fact detected. There are, however, only six such objects in the full map, and there does not seem to be a significant excess of these sources in the eastern part of the image. Admittedly, this latter point is difficult to assert with such a low number of detections.

We can constrain how many sources would need to be missing in the eastern population of disks by adding undetected sources (with undetected disks) to it. Using the same statistical tests as before, this will then give a lower limit to the number of undetected disk-bearing YSOs that would need to be added to the eastern population to make the two disk mass distributions indistinguishable. This happens if we assume that the eastern population is missing more than 70 infrared-identified YSOs that do not have detected disks at millimeter wavelengths either (i.e., for that number of added sources, $p< 0.05$). This means that there would be 30 nondetections for each detection without an IR counterpart in that part of the map. This ratio of detections to nondetections is far greater than what is expected for disks drawn from a Lupus-like population of $0.1 M_{\odot}$ stars~\citep{ansdell16,pascucci16}, and assumes that no sources are missing from the western population at all. Hence, any variations in the stellar mass sensitivity of the catalog between the eastern and western populations are not likely to dominate the difference in the two disk mass distributions. It does not mean, however, that such a bias is not present at all, and is an important caveat in the following discussion.

\subsection{Comparison to other star-forming regions}
\begin{figure*}[ht]
\begin{center}
                \includegraphics[width=\textwidth]{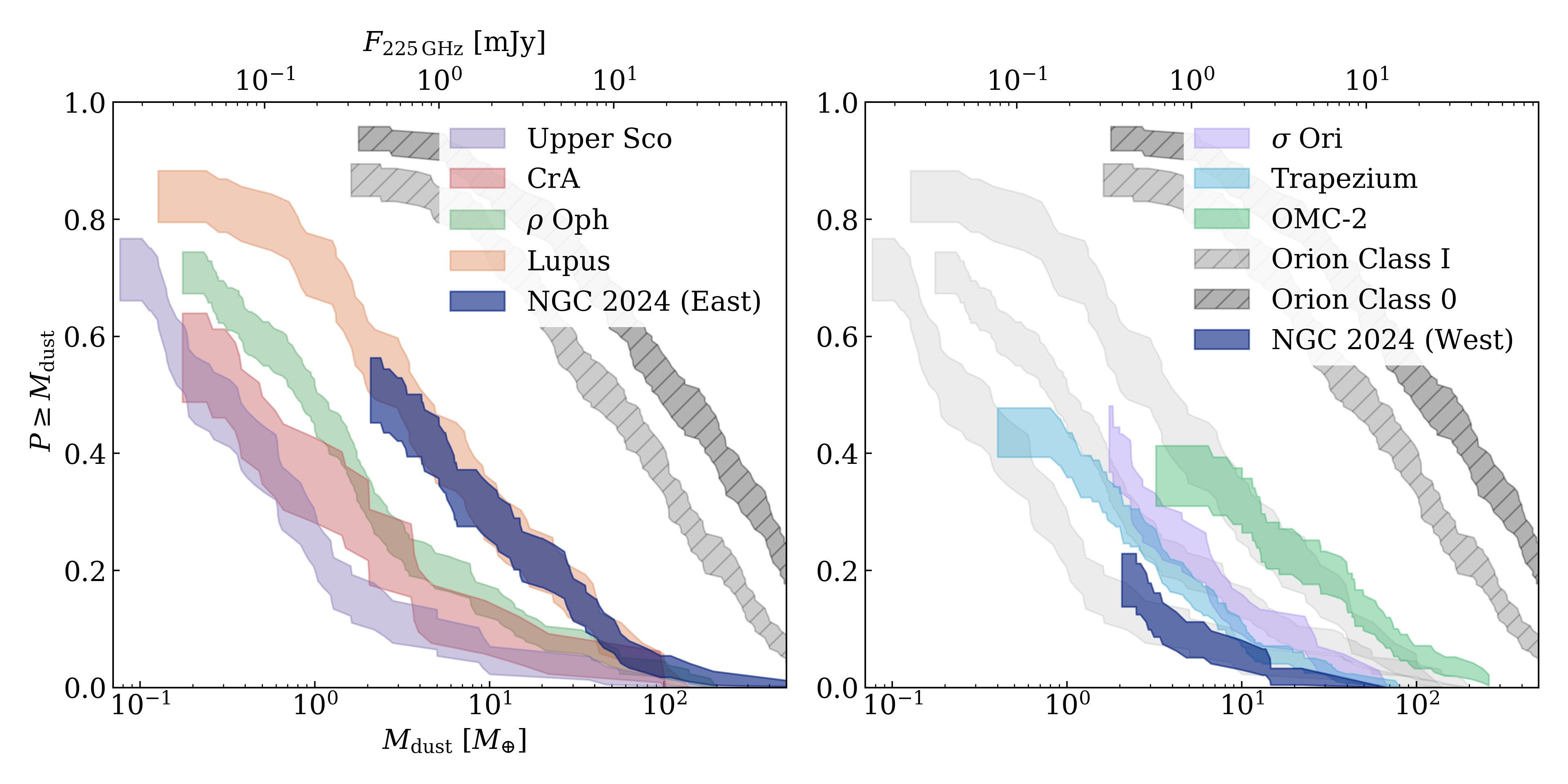}
                \caption{Disk mass distributions in NGC 2024 East (dark blue) (left) and West (dark blue) (right) compared to disk masses in other SFRs. Data for the low-mass regions was taken from~\citet{barenfeld16} for Upper Sco,~\citet{williams19} for $\rho$ Ophiuchi,~\citet{ansdell16} for Lupus,~and \citet{cazzoletti19} for Corona Australis. Disk masses across Orion are taken from~\citet{eisner18} for the Trapezium,~\citet{ansdell17} for $\sigma$ Ori, and~\citet{vanterwisga19} for OMC-2. Silhouettes in the right panel show the low-mass regions Lupus, $\rho$ Ophiuchi, and Upper Sco again, for comparison. The disk mass distributions for Class 0 and Class I objects across Orion from~\citet{tobin20} are shown in dark- and light gray in both panels, using the same temperature ($T_{\rm{eff}} = 20$\,K) and opacity as the more evolved regions.}
                \label{fig:compop}
\end{center}
\end{figure*}
In Fig.~\ref{fig:compop}, the disk mass distributions of the two subpopulations are compared to those of other SFRs. The left panel shows three well-studied, nearby, low-mass regions, where external photoevaporation is not expected to be a major factor in disk evolution. In contrast, the panel on the right shows Class II disk masses in different environments in Orion A and B: the Trapezium and $\sigma$ Orionis, where external photoevaporation has been shown to affect disk masses~\citep{mann14, ansdell17, eisner18}, and the OMC-2 region, which is to the north of the Trapezium and where no evidence for external photoevaporation has been found~\citep{vanterwisga19}. In this comparison, we cannot take into account possible differences in stellar masses, even though these may be significant, as discussed in Sect.~\ref{sec:mstar}, which is an important limitation. Similarly, the binarity fraction is important for disk masses, as binary systems have lower total disk masses in Taurus~\citep{akeson19,long19}. For many regions including NGC 2024, however, the number of binary YSOs is not well known, leading to additional uncertainty in the comparison of disk luminosities and masses. 

\subsubsection{NGC 2024 East}
The disk mass distributions of Lupus and NGC 2024 East resemble each other closely. Indeed, a log-rank test cannot formally distinguish between these populations. This is similarly true for a comparison with the disks in Taurus and Chamaeleon I~\citep{andrews13,pascucci16}. Within Orion, too, NGC 2024 East's disk mass distribution is indistinguishable from that in the OMC-2 cloud. All of these regions are typically given ages of 1--3\,Myr, older than NGC 2024, and older than the young stellar population in the cluster center.

In contrast, the approximately $\sim 2$\,Myr-old $\rho$ Oph star-forming region hosts a disk population that is markedly lower in mass than that of NGC 2024 East~\citep{cieza19, williams19}. However, $\rho$ Oph may also host two populations of YSOs of different ages, although this older population should be small ($<20\%$ of the full stellar population)~\citep{wilking05}. The disk mass distribution found in NGC 2024 East is also significantly more massive than that of the Corona Australis region's Class II population, despite most age estimates of this region indicating that it is young~\citep{cazzoletti19}.

Considering next the circumstellar dust masses of younger, embedded objects, Fig.~\ref{fig:compop} shows that the disk masses of NGC 2024 East lie significantly below those of Class I and Class 0 objects in the Orion A and B clouds~\citep{tobin20}, which have median disk masses of $>12\,M_{\oplus}$ in all cases, and would therefore be easily detected in a survey like this. In Serpens, disk masses of embedded sources are higher again than in Orion~\citep{tychoniec18}. This result is in line with the view that the evolution of disk masses does not proceed at a constant rate, and that Class 0 and Class I disks rapidly lose (millimeter-sized grain) luminosity, but do so only slowly once the Class II phase has been reached. In that case, the majority of solids evolution and planet formation occurs during the embedded phases of star formation, as suggested by the rich structures seen in HL Tau~\citep{hltau}, GY 91~\citep{sheehan18}, and other disks~\citep[e.g.,][]{long18, andrews18, vdmarel19}. This prediction can be tested by further observations: in this scenario, other disk properties, such as their continuum radii and gas fluxes, should also be similar in NGC 2024 and older regions.

\subsubsection{NGC 2024 West}
\label{sec:westvelsewhere}
The low masses of the western subpopulation of disks in this survey are remarkable, not just in contrast with the disks along the dense molecular ridge in the same region, but also when compared to ALMA observations of disk masses in other regions where external photoevaporation is important. Figure~\ref{fig:compop} shows NGC 2024 West next to three other star-forming regions in Orion, with different local radiation fields. In~\citet{vanterwisga19} it was shown that the OMC-2 population is apparently unaffected by external photoevaporation, due to its large distance from the Trapezium cluster. The oldest of the regions in this panel is $\sigma$ Ori at 3--5\,Myr. While the signature of external photoevaporation in this region is not as obvious relative to the effect of age, $\sigma$ Ori's ionizing star is cooler than those in NGC 2024 and the Trapezium, with a spectral type of O9. 

Comparing NGC 2024 West to the low-mass star-forming regions, it is most similar to Upper Sco, which has an age of 6--10\,Myr~\citep{barenfeld16}, and, interestingly, the previously-mentioned Corona Australis region~\citep{cazzoletti19}. While the age of NGC 2024 is inconsistent with the age estimates for Upper Sco, even if we allow for the significant uncertainties in stellar ages, the similarity to Corona Australis is more remarkable. As~\citet{cazzoletti19} discuss, it is likely that the low average mass in this region is a result of the initial conditions for disk formation there. We should therefore consider if the same initial conditions alone may explain the NGC 2024 West sample's disk masses. This seems to be unlikely: in Corona Australis, low disk masses are seen throughout the region, out to distances of approximately two parsec. In NGC 2024, however, the eastern subsample has much more massive disks despite these disks originating in a similar environment. More importantly, NGC 2024 does host sufficiently massive stars that the presence of at least some external irradiation should be considered.

Comparing NGC 2024 West to another region where external photoevaporation is present, we can conclude that its disks and those in the Trapezium cannot be distinguished by a log-rank test. NGC 2024 may in fact be less massive on average, although it would require deeper observations (or a larger sample) to test this conclusively. This result is in line with the expectation that stars outside the cluster core are older than those in the inner region (which make up our NGC 2024 East sample), and have similar ages to the disks in the core of the Trapezium~\citep{getman14} within the uncertainties and scatter in stellar age estimates. However, NGC 2024 West disks are on average at least 0.25\,pc from IRS 2b. In the Trapezium photoevaporation is much less important at this distance to $\theta^1$ Ori C, the primary source of ionizing radiation~\citep{mann14,eisner18}. As we suggested previously, the diminution of disk masses in NGC 2024 may be explained by the contribution of IRS 1 to the external photoevaporation. Alternatively, it is possible that disks in this population did start out with somewhat lower masses than disks in the Trapezium, in a similar way to what has been proposed for disks in $\rho$ Oph and R Corona Australis~\citep{williams19,cazzoletti19}. The older population extends over a larger area than the part observed here~\citep{getman14}. Thus, millimeter observations of disks in the outer part of the cluster will allow us to test this hypothesis.

\section{Conclusions}
In this article, we presented observations of a large field towards the center of NGC 2024, containing 179 protoplanetary disks, as well as several YSOs at earlier stages of evolution. The purpose of these observations was to characterize the disk masses in a young, massive star-forming region, and to use these masses to study how disk evolution is affected by the strong radiation fields and high stellar densities in such environments by comparing NGC 2024 to other regions. This comparison depends sensitively on our knowledge of the stellar populations of NGC 2024. By comparing CO velocity measurements with multi-wavelength surveys of YSOs and stellar ages, we can interpret the disk masses of the distinct populations in our field in a coherent way. This approach allows us to locate the Class II objects of NGC 2024 in the interstellar environment in much greater detail. In the future, a Guaranteed Time Observations program on the James Webb Space Telescope towards this region (ID 1190, PI: M. Meyer) will allow the properties of the stellar and substellar populations in this region to be constrained in greater detail, and to confirm the results presented here.

\begin{itemize}
        \item We observed 179 disks in NGC 2024 with ALMA in Band 6, of which $32 \pm 4\%$ are detected. Several other YSOs are detected, one of which is an X-ray source from which we detect a bright synchrotron flare.
        \item We identify two distinct populations of disks in NGC 2024: one in the eastern half of our field, centered on the dense molecular ridge hosting the FIR 1--5 sources, and one in the western half of the field.
        \item The eastern population is consistent with a $0.5$\,Myr age, has a significantly higher disk detection rate ($45 \pm 7\%$), and contains the most massive object in the sample, the IRC 101 transition disk. In terms of disk mass distributions, it resembles both the Lupus and the OMC-2 disk populations.
        \item The western population in contrast has much fainter disks, with a detection rate of $15 \pm 4\%$. This population is likely older ($\sim 1$\,Myr) and exposed to external FUV irradiation from IRS 2b and IRS 1, leading to rapid external photoevaporation.
\end{itemize}

%edited 11/09/2019
\begin{acknowledgements}
We would like to thank Tom Megeath for his insightful comments on the nature of infrared excess sources in Orion.

CFM acknowledges an ESO fellowship.

This project has received funding from the European Union's Horizon 2020 research and innovation programme under the Marie Sklodowska-Curie grant agreement No 823823 (DUSTBUSTERS).

This work was partly supported by the Deutsche Forschungs-Gemeinschaft (DFG, German Research Foundation) - Ref no. FOR 2634/1 TE 1024/1-1.
\end{acknowledgements}

%%% Bibliography goes here %%%
\bibliographystyle{aa}
\bibliography{ngc2024}

\begin{appendix}
\renewcommand\thefigure{\thesection.\arabic{figure}}
\renewcommand\thetable{\thesection.\arabic{table}}
\section{Cutouts of detected sources}
\label{app:cutouts}
\setcounter{figure}{0}
\raggedbottom

\begin{minipage}{0.96\textwidth}
        \centering
        \includegraphics[width=0.96\textwidth]{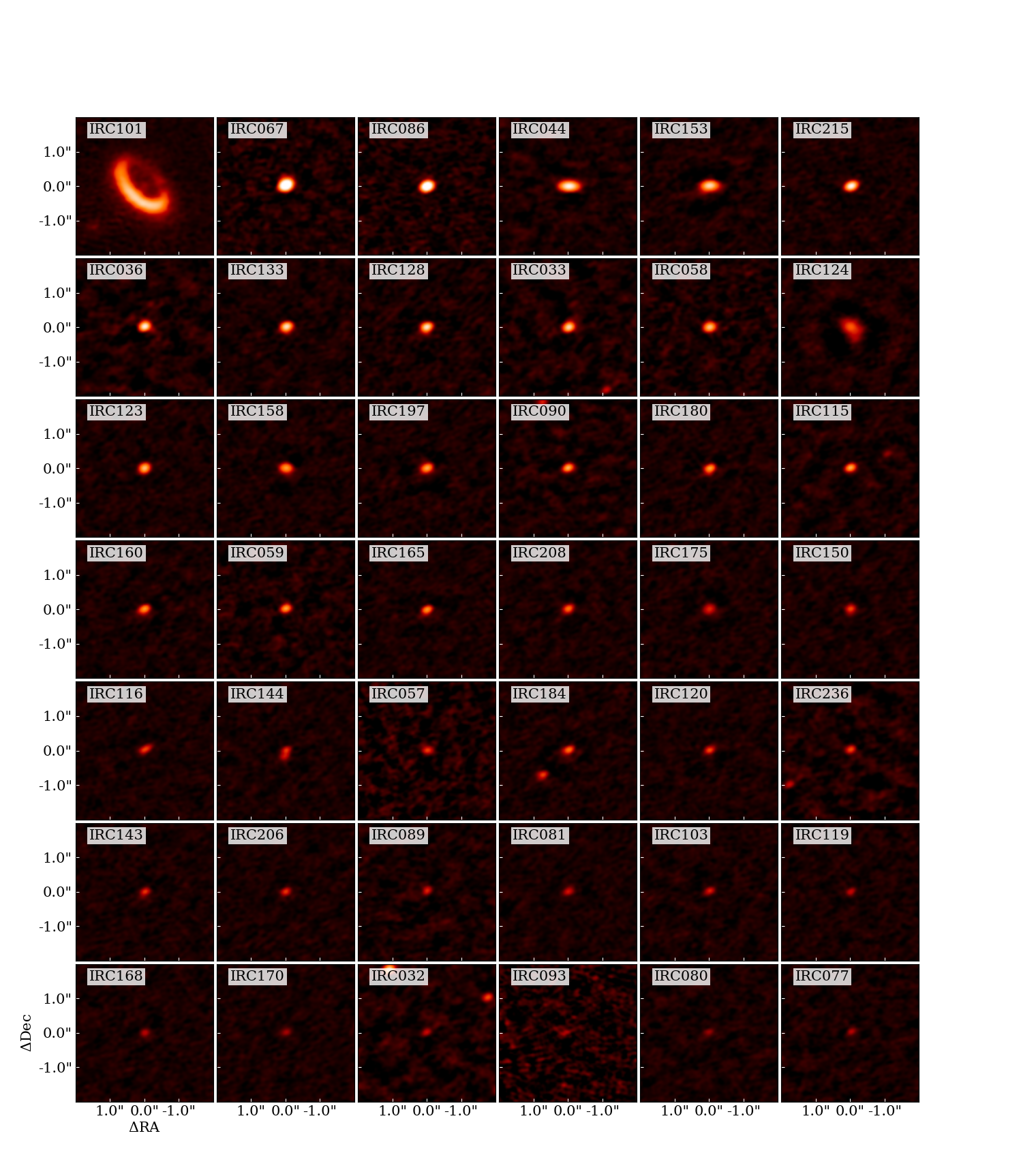}
        \captionof{figure}{Cutouts of the disks in the~\citet{meyer96} catalog detected in this study, sorted by integrated flux.}
        \label{fig:cutouts1}
\end{minipage}

\newpage
\onecolumn

\begin{minipage}{0.96\textwidth}
        \centering
        \includegraphics[width=0.96\textwidth, keepaspectratio]{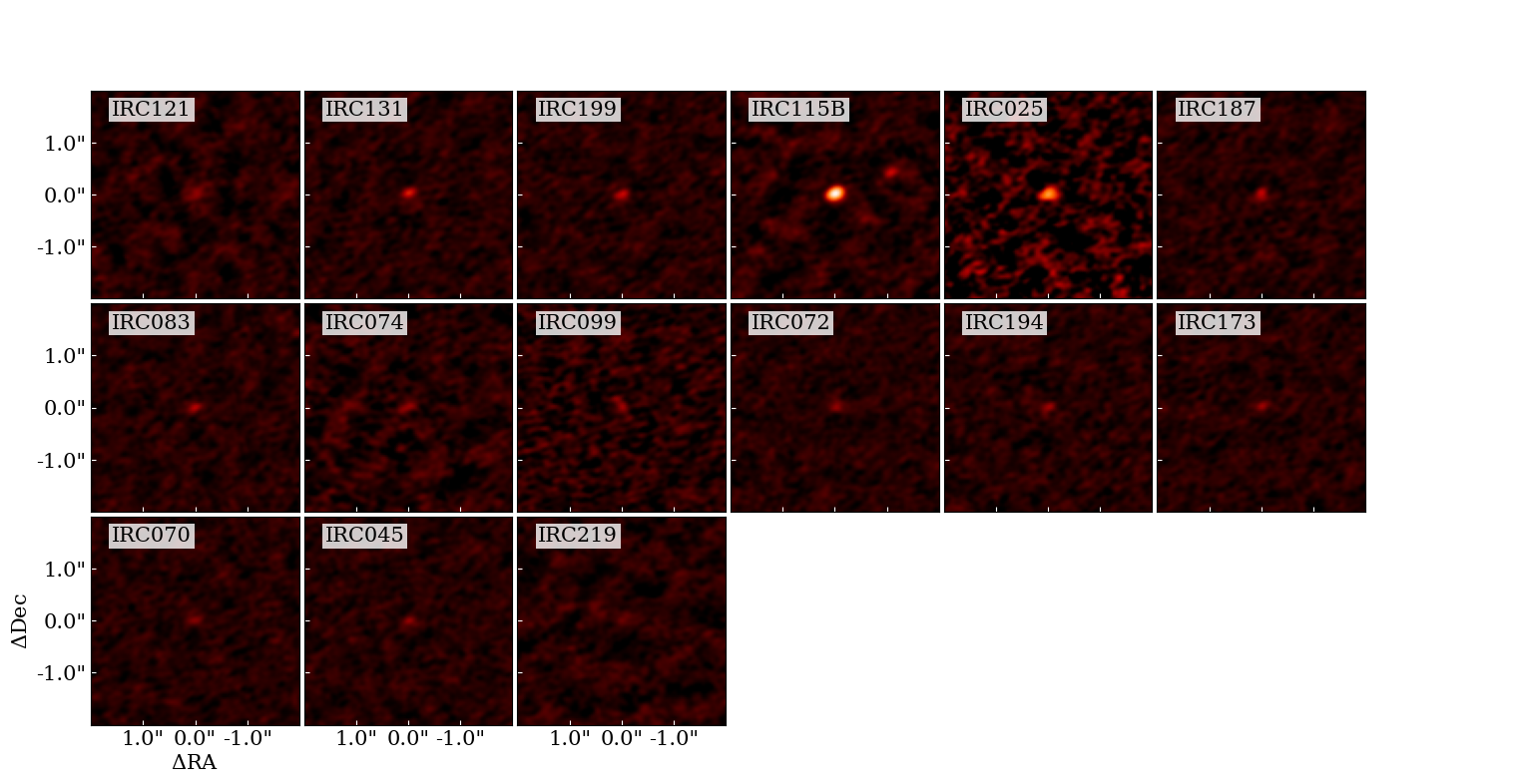}
        \captionof{figure}{Cutouts of the disks in the~\citet{meyer96} catalog detected in this study, sorted by integrated flux, continued from Figure~\ref{fig:cutouts1}.}
        \label{fig:cutouts2}
\end{minipage}

\begin{minipage}{0.96\textwidth}
        \centering
        \includegraphics[width=0.96\textwidth, keepaspectratio]{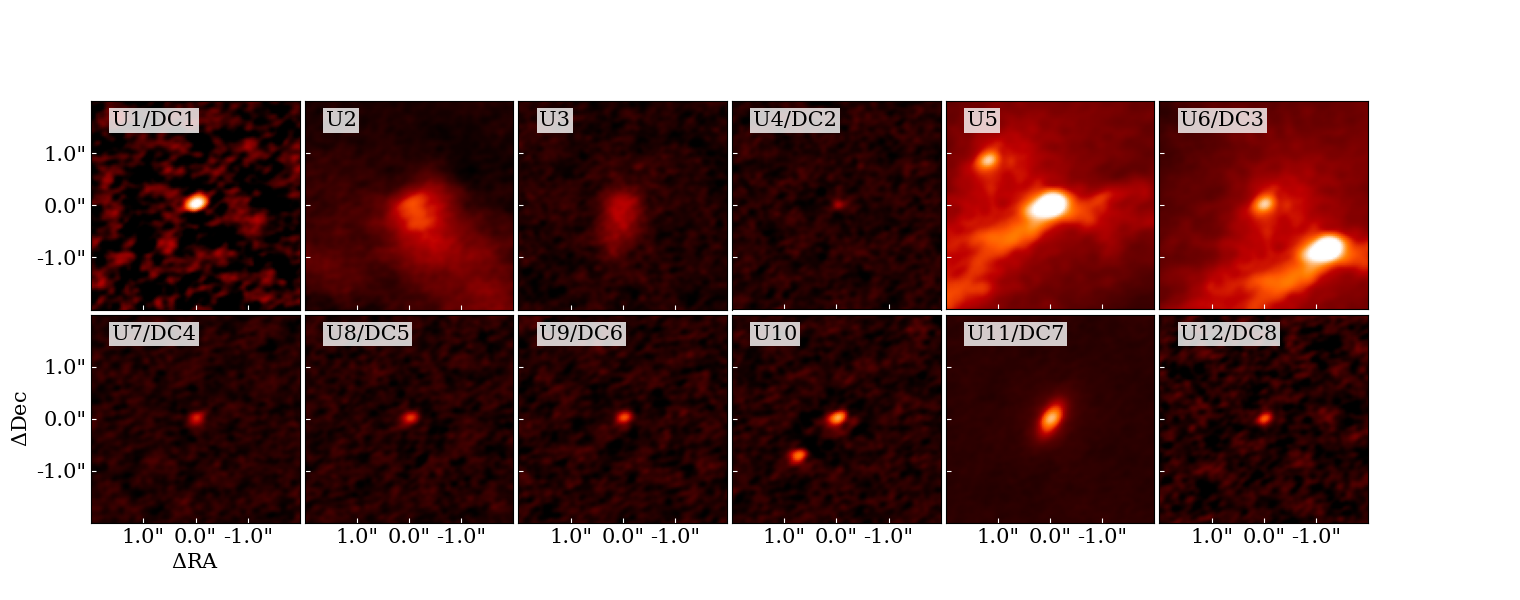}
        \captionof{figure}{Cutouts of detected sources without counterparts in the~\citet{meyer96} catalog detected in this study, sorted by integrated flux. Sources considered disk candidates are marked as DC 1--8.}
        \label{fig:cutouts3}
\end{minipage}

\vfill
\section{Flux and mass upper limits for undetected disks}
\label{app:uplimmass}
\renewcommand\thetable{\thesection.\arabic{table}}
\setcounter{table}{0}
\longtab[1]{
        \begin{longtable}{l l l l l}
                \caption{\label{tab:uplimmass} Continuum flux and mass upper limits (3$\sigma$) for the nondetected disks.} \\
                \hline\hline
                Name & RA & Dec & Flux & Mass \\
                &  &  & mJy & $M_{\oplus}$ \\
                \hline
                \endfirsthead
                \caption{(continued).}\\
                \hline\hline
                Name & RA & Dec & Flux & Mass \\
                &  &  & mJy & $M_{\oplus}$ \\
                \hline
                \endhead
                \hline
                \endfoot
                IRC156 & 5:41:46.24 & -1:53:48.0 & < 21.1 & < 108.5 \\
                IRC114 & 5:41:46.20 & -1:54:14.7 & < 7.9 & < 40.9 \\
                IRC167 & 5:41:46.18 & -1:53:42.5 & < 5.5 & < 28.4 \\
                IRC027 & 5:41:38.22 & -1:55:38.0 & < 4.3 & < 21.9 \\
                IRC053 & 5:41:46.04 & -1:55:02.2 & < 3.9 & < 20.0 \\
                IRC042 & 5:41:35.01 & -1:55:19.8 & < 3.6 & < 18.7 \\
                IRC115A & 5:41:35.03 & -1:53:29.3 & < 3.2 & < 16.6 \\
                IRC106 & 5:41:45.62 & -1:54:21.3 & < 2.9 & < 14.9 \\
                IRC161 & 5:41:45.98 & -1:53:45.4 & < 2.9 & < 14.7 \\
                IRC212 & 5:41:35.16 & -1:53:12.0 & < 2.7 & < 13.7 \\
                IRC122 & 5:41:45.91 & -1:54:12.6 & < 2.5 & < 13.0 \\
                IRC127 & 5:41:45.80 & -1:54:06.8 & < 1.6 & < 8.4 \\
                IRC102 & 5:41:45.17 & -1:54:22.8 & < 1.5 & < 7.5 \\
                IRC050 & 5:41:45.73 & -1:55:04.6 & < 1.4 & < 7.5 \\
                IRC063 & 5:41:45.72 & -1:54:49.6 & < 1.2 & < 6.4 \\
                IRC098 & 5:41:45.68 & -1:54:26.1 & < 1.2 & < 6.2 \\
                IRC113 & 5:41:35.39 & -1:54:19.4 & < 1.2 & < 6.0 \\
                IRC169 & 5:41:35.34 & -1:53:42.0 & < 1.2 & < 6.0 \\
                IRC190 & 5:41:35.43 & -1:53:31.8 & < 1.1 & < 5.9 \\
                IRC205 & 5:41:36.55 & -1:53:19.0 & < 1.1 & < 5.8 \\
                IRC188 & 5:41:43.94 & -1:53:29.2 & < 1.1 & < 5.6 \\
                IRC147 & 5:41:40.15 & -1:53:56.0 & < 1.1 & < 5.5 \\
                IRC155 & 5:41:35.45 & -1:53:50.8 & < 1.0 & < 5.3 \\
                IRC029 & 5:41:38.05 & -1:55:31.0 & < 1.0 & < 5.3 \\
                IRC139 & 5:41:45.42 & -1:53:57.8 & < 0.8 & < 4.3 \\
                IRC047 & 5:41:44.02 & -1:55:08.8 & < 0.8 & < 4.1 \\
                IRC176 & 5:41:45.36 & -1:53:34.5 & < 0.8 & < 4.1 \\
                IRC052 & 5:41:45.36 & -1:55:02.6 & < 0.7 & < 3.8 \\
                IRC162 & 5:41:35.72 & -1:53:47.7 & < 0.7 & < 3.4 \\
                IRC110 & 5:41:35.72 & -1:54:22.7 & < 0.7 & < 3.4 \\
                IRC073 & 5:41:44.90 & -1:54:42.5 & < 0.6 & < 3.1 \\
                IRC034 & 5:41:44.41 & -1:55:23.1 & < 0.6 & < 3.1 \\
                IRC177 & 5:41:35.85 & -1:53:38.2 & < 0.6 & < 3.0 \\
                IRC193 & 5:41:43.50 & -1:53:24.9 & < 0.5 & < 2.7 \\
                IRC040 & 5:41:44.17 & -1:55:19.8 & < 0.5 & < 2.7 \\
                IRC213 & 5:41:37.85 & -1:53:11.3 & < 0.5 & < 2.6 \\
                IRC117 & 5:41:43.16 & -1:54:13.9 & < 0.5 & < 2.6 \\
                IRC118 & 5:41:41.65 & -1:54:12.5 & < 0.5 & < 2.6 \\
                IRC092 & 5:41:42.66 & -1:54:30.1 & < 0.5 & < 2.5 \\
                IRC134 & 5:41:44.93 & -1:54:02.1 & < 0.5 & < 2.5 \\
                IRC209 & 5:41:41.84 & -1:53:13.1 & < 0.5 & < 2.5 \\
                IRC112 & 5:41:36.13 & -1:54:18.5 & < 0.5 & < 2.5 \\
                IRC225 & 5:41:36.25 & -1:53:33.5 & < 0.5 & < 2.5 \\
                IRC035 & 5:41:40.25 & -1:55:21.9 & < 0.5 & < 2.4 \\
                IRC210 & 5:41:37.34 & -1:53:13.5 & < 0.5 & < 2.4 \\
                IRC207 & 5:41:41.40 & -1:53:16.0 & < 0.5 & < 2.4 \\
                IRC204 & 5:41:38.54 & -1:53:17.7 & < 0.4 & < 2.3 \\
                IRC174 & 5:41:36.37 & -1:53:40.1 & < 0.4 & < 2.3 \\
                IRC186 & 5:41:38.25 & -1:53:33.7 & < 0.4 & < 2.3 \\
                IRC202 & 5:41:43.17 & -1:53:17.7 & < 0.4 & < 2.3 \\
                IRC107 & 5:41:44.72 & -1:54:19.9 & < 0.4 & < 2.3 \\
                IRC041 & 5:41:40.63 & -1:55:19.9 & < 0.4 & < 2.3 \\
                IRC189 & 5:41:40.05 & -1:53:29.7 & < 0.4 & < 2.3 \\
                IRC192 & 5:41:38.31 & -1:53:28.6 & < 0.4 & < 2.3 \\
                IRC164 & 5:41:36.38 & -1:53:48.5 & < 0.4 & < 2.3 \\
                IRC203 & 5:41:37.94 & -1:53:18.9 & < 0.4 & < 2.3 \\
                IRC195 & 5:41:39.47 & -1:53:26.6 & < 0.4 & < 2.3 \\
                IRC201 & 5:41:39.31 & -1:53:23.1 & < 0.4 & < 2.3 \\
                IRC198 & 5:41:38.15 & -1:53:24.2 & < 0.4 & < 2.3 \\
                IRC231 & 5:41:44.72 & -1:54:32.0 & < 0.4 & < 2.3 \\
                IRC200 & 5:41:38.69 & -1:53:23.5 & < 0.4 & < 2.3 \\
                IRC224 & 5:41:36.76 & -1:53:36.4 & < 0.4 & < 2.3 \\
                IRC039 & 5:41:38.21 & -1:55:19.6 & < 0.4 & < 2.3 \\
                IRC163 & 5:41:37.57 & -1:53:47.9 & < 0.4 & < 2.2 \\
                IRC183 & 5:41:39.55 & -1:53:35.4 & < 0.4 & < 2.2 \\
                IRC138 & 5:41:44.55 & -1:53:58.5 & < 0.4 & < 2.2 \\
                IRC152 & 5:41:37.82 & -1:53:51.1 & < 0.4 & < 2.2 \\
                IRC178 & 5:41:40.04 & -1:53:36.9 & < 0.4 & < 2.2 \\
                IRC191 & 5:41:41.41 & -1:53:28.3 & < 0.4 & < 2.2 \\
                IRC159 & 5:41:36.54 & -1:53:50.3 & < 0.4 & < 2.2 \\
                IRC182 & 5:41:41.12 & -1:53:32.8 & < 0.4 & < 2.2 \\
                IRC172 & 5:41:44.21 & -1:53:37.9 & < 0.4 & < 2.2 \\
                IRC146 & 5:41:38.10 & -1:53:56.9 & < 0.4 & < 2.2 \\
                IRC171 & 5:41:40.00 & -1:53:40.5 & < 0.4 & < 2.2 \\
                IRC148 & 5:41:39.02 & -1:53:54.6 & < 0.4 & < 2.2 \\
                IRC149 & 5:41:36.63 & -1:53:56.2 & < 0.4 & < 2.2 \\
                IRC181 & 5:41:43.32 & -1:53:32.5 & < 0.4 & < 2.2 \\
                IRC157 & 5:41:44.10 & -1:53:46.3 & < 0.4 & < 2.2 \\
                IRC145 & 5:41:39.05 & -1:53:58.2 & < 0.4 & < 2.2 \\
                IRC135 & 5:41:36.46 & -1:54:04.5 & < 0.4 & < 2.2 \\
                IRC185 & 5:41:42.11 & -1:53:32.1 & < 0.4 & < 2.2 \\
                IRC151 & 5:41:41.47 & -1:53:52.8 & < 0.4 & < 2.2 \\
                IRC179 & 5:41:41.71 & -1:53:34.7 & < 0.4 & < 2.2 \\
                IRC166 & 5:41:41.96 & -1:53:44.9 & < 0.4 & < 2.2 \\
                IRC141 & 5:41:41.45 & -1:53:57.9 & < 0.4 & < 2.2 \\
                IRC136 & 5:41:39.33 & -1:54:01.8 & < 0.4 & < 2.2 \\
                IRC130 & 5:41:36.72 & -1:54:07.5 & < 0.4 & < 2.2 \\
                IRC126 & 5:41:39.00 & -1:54:09.0 & < 0.4 & < 2.2 \\
                IRC132 & 5:41:43.97 & -1:54:02.6 & < 0.4 & < 2.2 \\
                IRC140 & 5:41:43.13 & -1:53:56.0 & < 0.4 & < 2.2 \\
                IRC111 & 5:41:37.47 & -1:54:18.7 & < 0.4 & < 2.1 \\
                IRC129 & 5:41:40.20 & -1:54:07.5 & < 0.4 & < 2.1 \\
                IRC223 & 5:41:38.33 & -1:54:19.9 & < 0.4 & < 2.1 \\
                IRC125 & 5:41:40.93 & -1:54:08.2 & < 0.4 & < 2.1 \\
                IRC108 & 5:41:44.31 & -1:54:20.4 & < 0.4 & < 2.1 \\
                IRC082 & 5:41:36.51 & -1:54:41.5 & < 0.4 & < 2.1 \\
                IRC104 & 5:41:38.39 & -1:54:26.0 & < 0.4 & < 2.1 \\
                IRC222 & 5:41:41.88 & -1:54:20.8 & < 0.4 & < 2.1 \\
                IRC097 & 5:41:38.77 & -1:54:28.3 & < 0.4 & < 2.1 \\
                IRC226 & 5:41:37.61 & -1:54:25.5 & < 0.4 & < 2.1 \\
                IRC054 & 5:41:42.46 & -1:55:02.8 & < 0.4 & < 2.1 \\
                IRC043 & 5:41:41.46 & -1:55:13.4 & < 0.4 & < 2.1 \\
                IRC051 & 5:41:42.14 & -1:55:05.8 & < 0.4 & < 2.1 \\
                IRC230 & 5:41:42.37 & -1:54:59.3 & < 0.4 & < 2.1 \\
                IRC069 & 5:41:42.59 & -1:54:47.2 & < 0.4 & < 2.1 \\
                IRC100 & 5:41:37.81 & -1:54:28.1 & < 0.4 & < 2.1 \\
                IRC055 & 5:41:41.81 & -1:55:01.6 & < 0.4 & < 2.1 \\
                IRC087 & 5:41:43.97 & -1:54:37.7 & < 0.4 & < 2.1 \\
                IRC216 & 5:41:43.69 & -1:54:37.1 & < 0.4 & < 2.1 \\
                IRC075 & 5:41:42.75 & -1:54:43.1 & < 0.4 & < 2.1 \\
                IRC076 & 5:41:43.65 & -1:54:39.7 & < 0.4 & < 2.1 \\
                IRC084 & 5:41:41.36 & -1:54:38.0 & < 0.4 & < 2.1 \\
                IRC088 & 5:41:42.78 & -1:54:37.2 & < 0.4 & < 2.1 \\
                IRC091 & 5:41:38.09 & -1:54:33.7 & < 0.4 & < 2.1 \\
                IRC094 & 5:41:37.90 & -1:54:32.8 & < 0.4 & < 2.1 \\
                IRC079 & 5:41:41.60 & -1:54:41.3 & < 0.4 & < 2.1 \\
                IRC214 & 5:41:37.71 & -1:54:33.5 & < 0.4 & < 2.1 \\
                IRC048 & 5:41:37.81 & -1:55:09.4 & < 0.4 & < 2.1 \\
                IRC061 & 5:41:38.19 & -1:54:55.2 & < 0.4 & < 2.1 \\
                IRC078 & 5:41:39.50 & -1:54:40.9 & < 0.4 & < 2.1 \\
                IRC233 & 5:41:37.76 & -1:54:39.0 & < 0.4 & < 2.1 \\
                IRC056 & 5:41:39.00 & -1:55:01.3 & < 0.4 & < 2.1 \\
        \end{longtable}
}

\end{appendix}

\end{document}